\documentclass[10pt,preprint]{aastex}

\shorttitle{Photometry of Common Proper Motion Binaries}
\shortauthors{Li et al.}
\newcommand{\degree}{\ensuremath{^\circ}}

\begin{document}

\title{Optical $BVRI$ Photometry of Common Proper Motion F/G/K+M Wide Separation Binaries}

\author{Ting Li\altaffilmark{1,2}, Jennifer Marshall\altaffilmark{1,2,3}, S{\'e}bastien L{\'e}pine\altaffilmark{4,5,7}, Patrick Williams\altaffilmark{1}, Joy Chavez\altaffilmark{1,7}}

\altaffiltext{1}{Department of Physics \& Astronomy, Texas A \& M University, College Station, TX 77840, USA}
\altaffiltext{2}{George P. and Cynthia Woods Mitchell Institute for Fundamental Physics and Astronomy, College Station, TX, 77843-4242, USA}
\altaffiltext{3}{The Observatories of the Carnegie Institute of Washington, 813 Santa Barbara Street, Pasadena, CA 91101, USA}
\altaffiltext{4}{Department of Physics \& Astronomy, Georgia State University, Atlanta, GA 30302-4106, USA}
\altaffiltext{5}{Department of Astrophysics, American Museum of Natural History, New York, NY 10024, USA}
\altaffiltext{6}{Department of Physics, Graduate Center, City University of New York, New York, NY 10016, USA}
\altaffiltext{7}{Gemini Observatory, 670 N. A’ohoku Pl., Hilo, Hawai’i 96720-2700, USA}

\begin{abstract}
We present optical ($BVRI$) photometric measurements of a sample of 76 common proper motion wide separation main sequence binary pairs. The pairs are composed of a F-, G-, or K-type primary star and an M-type secondary. The sample is selected from the revised NLTT catalog and the LSPM catalog. The photometry is generally precise to 0.03 mag in all bands. We separate our sample into two groups, dwarf candidates and subdwarf candidates, using the reduced proper motion (RPM) diagram constructed with our improved photometry. The M subdwarf candidates in general have larger $V-R$ colors than the M dwarf candidates at a given $V-I$ color. This is consistent with an average metallicity difference between the two groups, as predicted by the PHOENIX/BT-Settl models. The improved photometry will be used as input into a technique to determine the metallicities of the M-type stars. 
\end{abstract}


\keywords{Galaxy:halo---Galaxy:solar neighborhood---stars:subdwarfs---stars:late-type}


\section{Introduction}\label{sec:intro}

M-type main sequence stars are small and faint and are therefore difficult to study, but they are the most numerous stars in the Galaxy and so form an important subset of the Galactic population. Late-type members of the Galactic halo, M subdwarfs, could in principle give the most complete census of the distributions of both metallicities and kinematics of the local halo and could be used to investigate the chemical enrichment history and merger history of the Galaxy.  In addition, M dwarfs have become the target for many exoplanet surveys, since low-mass Earth- or Neptune-size planets orbiting lower mass stars are more easily detected with the Doppler or transit techniques than those orbiting higher mass stars. An improved understanding of M dwarf stellar atmospheres will lead to a better understanding of these exoplanet hosts.

Despite the potential importance of late-type main sequence stars on Galactic and exoplanet studies, these low mass stars remain one of the least understood stellar types. The spectra of M-type stars are complicated with many broad molecular bands due to abundant diatomic and triatomic molecules such as TiO, VO, H$_2$0, CO and CaH \citep{Gizis1997}.  Historically, atmospheric models have had difficulties fitting these broad features with adequate precision to allow for accurate metallicity estimates \citep{Bonfils2005}. Traditional methods of determining abundance from high-resolution spectra of weak atomic lines used for hotter stars are inaccurate for cooler M-type stars since the local pseudo-continuum estimated from a high-resolution spectrum is polluted by a forest of molecular absorption features. Moreover, the atmospheric models of the M-type stars still cannot reproduce the details of high-resolution spectra due to the limitation of the molecular opacity database. So far, deriving metallicity from high resolution spectra has only been attempted for a small sample of stars \citep[][]{Woolf2005,Bean2006,Onehag2012} since such measurements are difficult.

Progress has recently been made to measure the metallicities of M-type main sequence stars using visual binary pairs that contain both a solar type (F-, G-, or K-type) primary and an M-type companion. Since both components of a binary pair presumably formed from the same molecular cloud, the metallicity of an M-type companion can be assumed to be the same as that of the F-, G-, K-type primary. One therefore can create an empirical calibration to determine metallicities of M stars using the observable features of the M-type secondary and the metallicities of the primary stars derived from high-resolution spectra. \cite{Bonfils2005} pioneered the use of F/G/K+M binaries to obtain a photometric calibration of M dwarf metallicities with the closest and brightest M dwarfs that have known parallaxes. \cite{Woolf2006a} used molecular line indices (e.g. TiO, CaH) at visible wavelengths in moderate-resolution spectra as a proxy for metallicity.  In the past few years, a few studies have used the spectral lines and spectral indices for metallicity sensitive features in the near-infrared spectra \citep[][]{RojasAyala2012,Mann2013,Newton2014}. However, most of this work has been applied to stars with near-solar metallicity ($\rm{[Fe/H]} > -1$). Little work has been done to date on measuring the metallicities of M subdwarfs \citep[$\rm{[Fe/H]} < -1$; see, however,][]{Woolf2009}. 

We plan to develop an empirical calibration of M subdwarf metallicities similar to the work of \cite{Woolf2006a} but over a wider range of metallicities ($-2.5 < \rm{[Fe/H]} < -0.5$). We have selected a large sample of common proper motion binaries having kinematics consistent with the Galactic halo population and we will measure the metallicities of the primaries using standard techniques. Accurate photometry is necessary for deriving the atmospheric parameters of the primaries.  Line indices associated with TiO and CaH features measured in moderate-resolution spectroscopy of the secondaries will be used to calibrate their metallicities and, again, improved photometry of the secondaries will be essential in establishing the metallicity calibration. 

In this paper we present the results of the $BVRI$ photometry of the selected common proper motion F/G/K+M main sequence binaries. Throughout this paper, we will use the term ``M dwarf" to refer to M-type main sequence stars that, based on their colors and kinematics, are likely members of the Galactic disk population (and are therefore likely to be relatively metal-rich, i.e., $-0.5 < \rm{[Fe/H]} < +0.5$).  We will also refer to ``M subdwarfs," the generally more metal-poor ($\rm{[Fe/H]} < -0.5$) analogs of main sequence M dwarfs, that are likely to belong to the Galactic halo or thick disk populations.  Section 2 briefly describes the sample selection. This is followed by a description of the observations and data reduction in Section 3. In Section 4, we present the photometry and errors of the sample. An improved RPM diagram and a new color-color diagram for this sample are also derived.  In a future paper, spectroscopic observations of these binaries together with the photometry presented in this paper will be used to derive an empirical calibration of metallicities for M subdwarfs. (Marshall et al. 2014, in preparation)

\section{Sample Selection}\label{sec:sample}

We selected stars from a preliminary version of the LSPM-South (L{\'e}pine, private communication) and LSPM-North catalog \citep{Lepine2005}; hereafter we will refer to the combination of these two catalogs as the LSPM catalog. The LSPM catalog is from the SUPERBLINK proper motion survey based on the analysis of scans from the Digitized Sky Survey \citep{Lepine2005b,Lepine2008}. We also selected ``halo" binaries from the published list of wide binaries from the revised New Luyten Two-Tenths Catalog (rNLTT) by \cite{Chaname2004}. The rNLTT catalog contains astrometry and optical/infrared photometry for the vast majority of NLTT stars ($\mu > 0.18$\arcsec$/yr$) lying in the overlap of regions covered by POSS I and by the second incremental Two Micron All Sky Survey (2MASS) release \citep{Gould2003,Salim2003}. The infrared (primarily $J$ band) photometry of both catalogs comes from 2MASS \citep{Skrutskie2006}, which is well calibrated and has good photometric precision, with errors typically less than 3\%. The quality of the optical photometry is not as good. Both catalogs adopted $V$ photometry from the Tycho-2 Catalog \citep{Hog2000} for most stars with $V<12$ mag and the photometic errors are about 0.10 mag. For  $V>12$ mag, $V$ photometry of LSPM catalog is converted from the photographic magnitudes of the blue and red plates from the USNO-B1.0 catalog \citep{monet2003}, while similar derivation is done for the photometry in the rNLTT catalog but the photographic magnitude is from the USNO-A catalog \citep[][]{monet1996, monet1998}. In both catalogs, the photometric errors from blue and red plate are about 0.3 mag. Binaries from both catalogs were selected on the basis of a given pair of stars having a separation of less than 900$\arcsec$ and larger than 3$\arcsec$ and having common proper motions. The pairs are generally required to have a proper motion difference $\Delta\mu\ < 20$ mas/yr. 

The sample was then selected to contain subdwarfs via their placement on a reduced proper motion (RPM) diagram, which plots $H_m=m+5\log(\mu)+5$ versus color and is used to separate stars into distinct kinematic populations \citep{Marshall2007,Lepine2007}.  The RPM diagram uses the proper motion as a distance proxy and is thereby similar to a Color-Magnitude Diagram (CMD), in which subdwarfs have bluer colors than main sequence dwarfs at a given luminosity due to reduced metal opacity in the optical. More importantly, stellar members of the local halo population may have large proper motions, which increases the value of $H_m$ relative to local disk stars.  Because of these two effects, an RPM diagram that is constructed using high precision photometry can be used to separate the dwarf (disk) and subdwarf (halo) kinematic populations cleanly and to search efficiently for halo subdwarf candidates and even extreme subdwarfs ($\rm{[Fe/H]} < -2$) candidates \citep{Marshall2008a}. 

\begin{figure}[htbp]
  \epsscale{1}
  \plotone{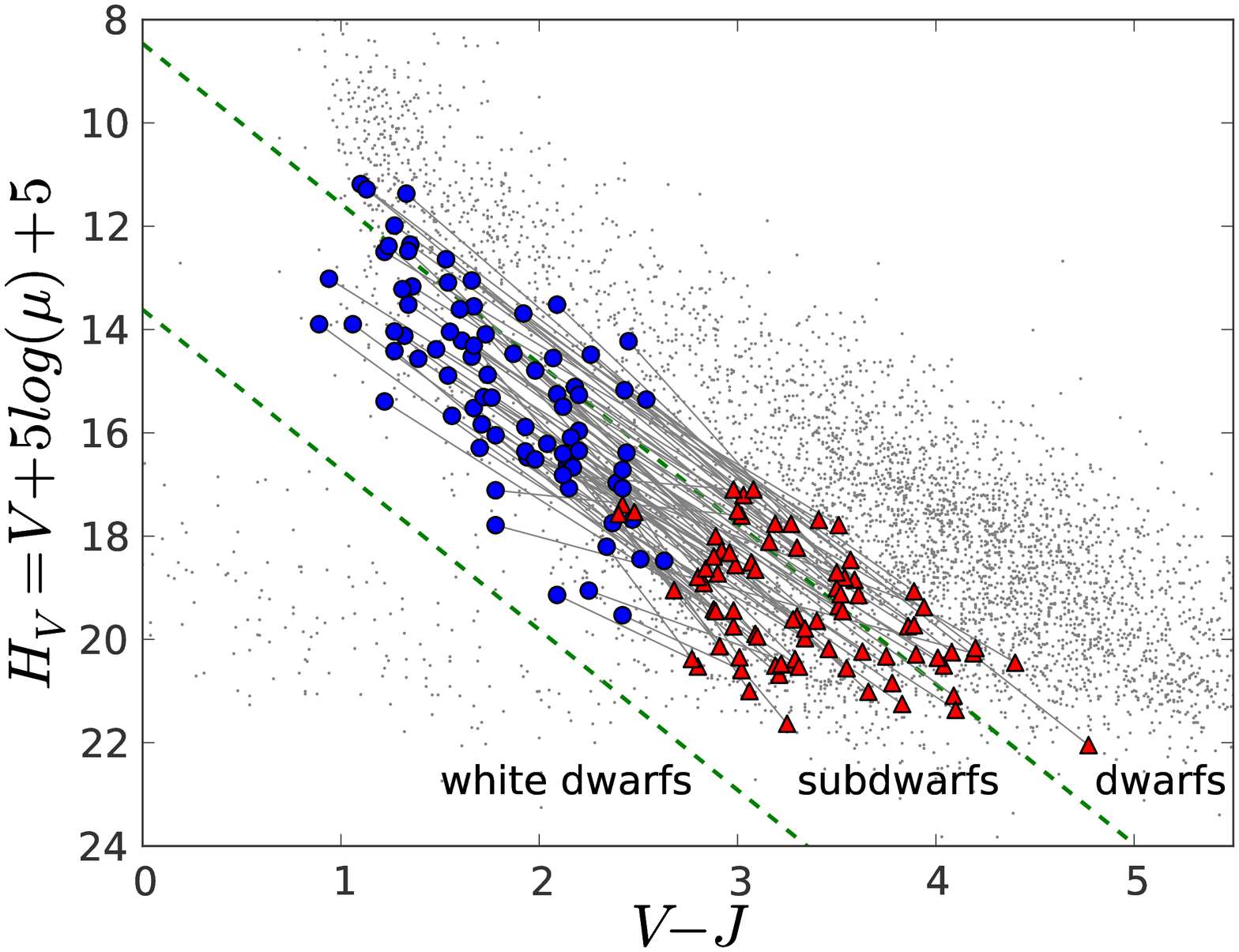}
  \caption{Reduced proper motion (RPM) diagram of the sample pairs. Blue filled circles indicates F/G/K-type candidate subdwarf primaries and red filled triangles represent the M-type candidate subdwarf secondaries. Lines connect the primary and secondary of each pair. Discriminator lines separating the solar metallicity dwarfs, metal-poor subdwarfs, and white dwarfs are drawn at $\eta=$ 0 and $\eta=$ 5.15, as defined by \cite{Salim2003}. Overplotted grey dots are stars in the LSPM-North catalog. \citep{Lepine2005} }
  \label{fig:rpm_old}
\end{figure}

Figure \ref{fig:rpm_old} shows the RPM diagram of the sample from which these binaries were drawn; lines connect the primary and secondary to show that they are very likely to have similar metallicities and the same proper motion. Two discriminator lines with $\eta=0$ and $\eta=5.15$ are drawn on the plot to separate the Main-Sequence (MS), subdwarfs (SD) and white dwarfs (WD), as defined by \cite{Salim2003}:

\begin{equation}\label{eq:rpmeq}
\eta(H_V,V-J,\sin b)=H_V-3.1(V-J)-1.47\left | \sin b \right |-7.73
\end{equation}
where b is the Galactic latitude of the targets. The discriminator lines presented here are drawn for a sample of low Galactic latitude $b=\pm30^{\circ}$. Note that, due to the large errors of the previously published photometry, we made a relatively generous choice of discriminator ($\eta > -1.5$) in order to include all the potential subdwarf candidates.  

The final requirement is a color cut so that the primaries generally have $(V-J) < 2.5$ and the secondaries have $(V-J) > 2.8$; in this way we select primaries that are likely to be F, G, or K subdwarfs and secondaries that are M subdwarfs. These criteria yielded a list of 9 pairs from \cite{Chaname2004} and 65 pairs from LSPM catalog, which are observable from the southern hemisphere ($\delta$ $<$ +20$\degree$). Two pairs are both in \cite{Chaname2004} and LSPM catalog; for these two pairs we adopt the information from LSPM catalog. Given the chance that the photometric errors in the catalog were higher than expected, we also included 3 pairs with primaries having $(V-J) > 2.5$ and 5 pairs with the secondaries having $(V-J) < 2.8$. We therefore have a total sample of 80 pairs in this work; the position, proper motion, $V$ magnitude, $(V-J)$ color, and separation of these candidate pairs are listed in Table \ref{table:sample}. The secondaries of three pairs (PM I00329+1805, PM I04327+0820, PM I06436+0851 as primaries) were identified in the preliminary version of the LSPM-South catalog but they are not included in the final version of the catalog, because they are marginally detected on Palomar Sky Survey plates. However our observations have indeed confirmed that these three pairs are co-moving objects, so they are included in this study. In Table \ref{table:sample}, the secondaries of these three pairs are named with the same name as their primaries with a ``-2'' appended.

\section{Observation and Data Reduction}\label{sec:data}
Absolute photometry was obtained at Las Campanas Observatory on the 1-meter Henrietta Swope Telescope on two observing runs in February and September 2008. The SITe$\#3$ CCD detector (0$\farcs$435/pixel) with a standard $BVRI$ Johnson/Kron-Cousins filter set was used for both runs. Observations in $BVRI$ filters were obtained for each candidate. Optical photometric measurements of 76 of the 80 candidate pairs in the sample were obtained under photometric conditions during 4 nights in February and 3 nights in September 2008. About half of the candidates have multiple observations that were obtained as a consistency check of the photometry. Since the primaries are generally 3-5 magnitude brighter than the secondaries, one long (100-300 seconds) exposure and one short (3-20 seconds) exposure were taken for most pairs, so as to achieve Signal-to-Noise Ratio (S/N) $>$ 30 for both primary \& secondary stars. Landolt standard stars \citep{Landolt1992} were used to calibrate the photometry. A Landolt field was observed approximately once per hour on each photometric night. Care was taken to ensure the standard fields were observed at a wide range of air masses. Standard fields were selected to contain stars with a wide range of colors; in particular, many standard stars were observed with very red colors in order to better calibrate the M subdwarfs. It should be noted, however, that photometric calibration of very red stars using this technique is difficult due to the small number of red stars in Landolt's catalog. For calibration purposes, both dome and twilight-sky flat fields were constructed each day. 

The data were reduced with IRAF\footnote{IRAF is distributed by the National Optical Astronomy Observatory, which is operated by the Association of Universities for Research in Astronomy, Inc., under cooperative agreement with the National Science Foundation} using standard CCD data reduction techniques. Each science frame was first bias-subtracted and linearity corrected and a daily flat field was applied. The subtraction of a bias level is determined from the overscan columns.  

\begin{figure}[htbp]
  \epsscale{1}
  \plotone{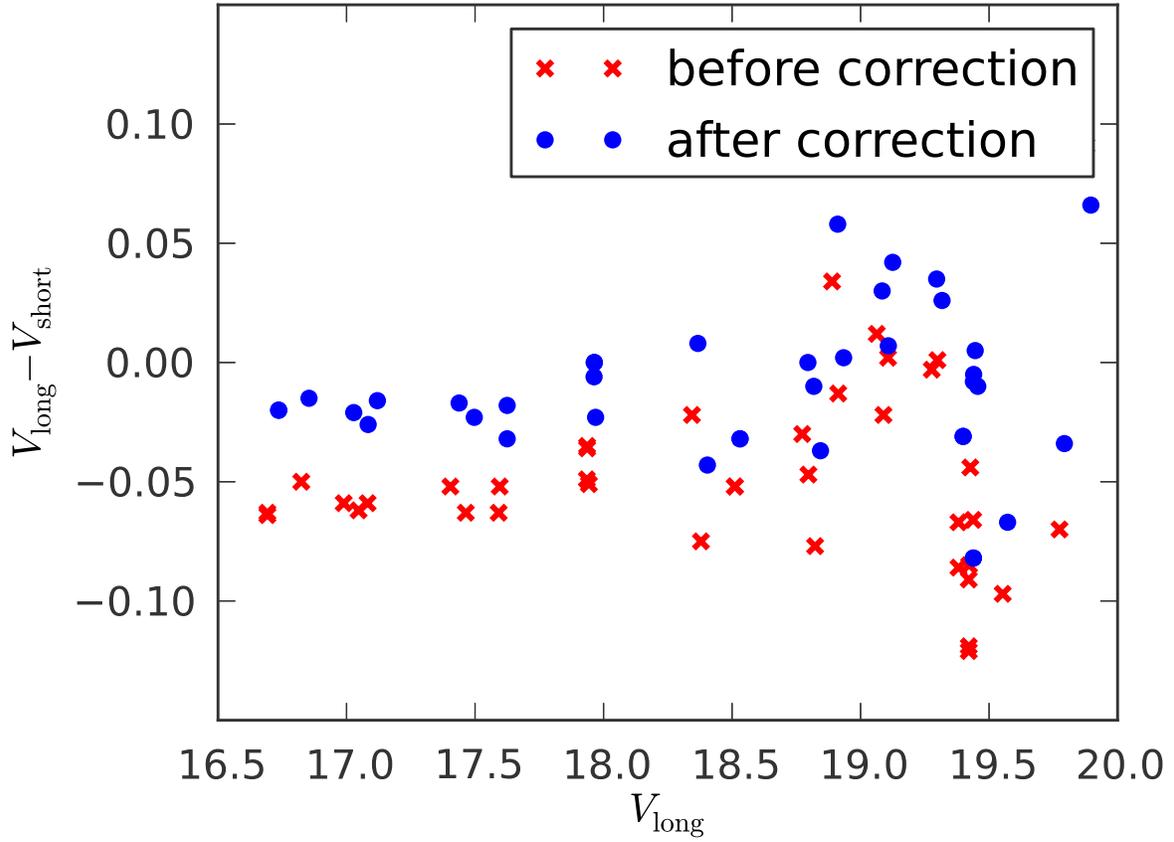}
  \caption{Comparison of the instrumental magnitudes in $V$ band of the same stars in two consecutive frames with different exposure times, 300 seconds and 30 second, before and after a linearity correction is applied. The red cross markers indicate that the difference is as large as 0.06 magnitude before the linearity correction. The blue filled circles show that the difference is about 0.01 mag after the linearity correction.}
  \label{fig:dif_lin}
\end{figure}

\begin{figure}[htbp]
  \epsscale{1}
  \plotone{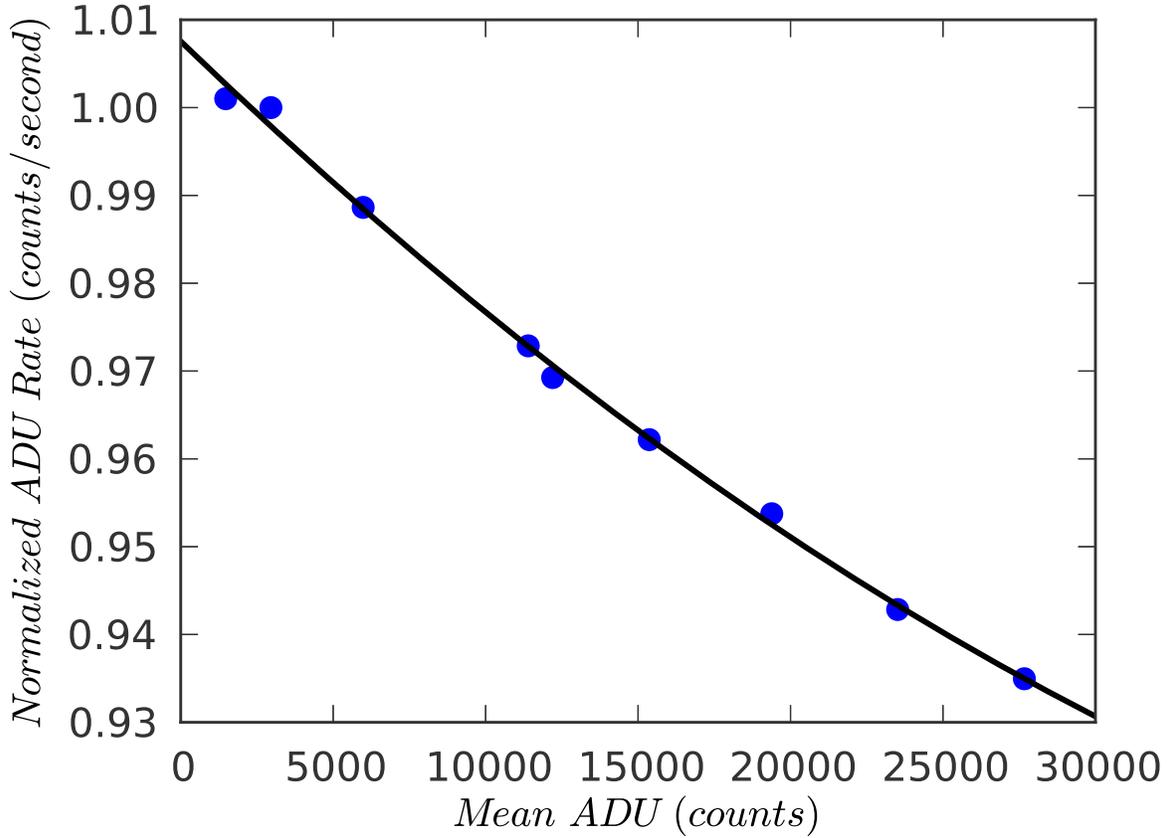}
  \caption{Linearity tests were performed using a sequence of images taken with increasing exposure times. The analog to digital unit (ADU) rate, counts per second, was normalized to the neighboring 8-second reference frame to compensate for the instability of the quartz lamp during the test. The normalized ADU rate is not a constant over all frames, which shows a non-linearity problem for this CCD. We therefore fit the data to a second-order polynomial function and use that to correct each frame. }
  \label{fig:lin_fit}
\end{figure}

We measure the non-linearity of the detector by examining the instrumental magnitude of stars in two consecutive frames with different exposure times since we took shorter exposures for the primaries and longer exposures for the secondaries. The instrumental magnitudes of random stars in the same field measured from the longer exposure frame (300 seconds) were about 0.06 mag brighter than that from the shorter exposure frame (30 seconds), as shown in Figure \ref{fig:dif_lin}. Therefore, a linearity correction was made for each frame using the IRAF utility \emph{irlincor} with a conversion derived from a linearity test constructed once per observing run. The linearity test consists of a sequence of images taken with increasing exposure times (4, 8, 16, 8, 32, 8, 64, 8... seconds). A reference frame with 8 seconds was taken for every other exposure to compensate for the instability of the quartz lamp during the linearity test. Then the mean analog to digital units (or ADU) of each frame was computed and the ADU rate (i.e. counts per second) was normalized to the neighboring 8-second reference frame. Figure \ref{fig:lin_fit} shows this normalized ADU rate as a function of the mean ADU of each frame. The data are well fit by a  second-order polynomial, with 3 fitting coefficients as the inputs to the IRAF command \emph{irlincor}. However, there are still slight systematic errors of $\approx$0.015 mag in the bright stars, possibly due to the fact we do not have any linearity information below 2000 ADU. The linearity correction as described above was applied to each frame after the bias subtraction but before the flat field correction.

We also investigated the precision of shutter timing for the system. The instrument was commanded  to take a 1 second exposure; actually an exposure of 2.13-2.19 seconds was taken (as noted in the image header). We therefore only use images with exposures longer than 3 seconds in our analysis. We also examined the shutter timing effects on the shutter-correction frames constructed once per run: the shutter-correction frames were constructed with multiple exposures without reading each exposure out (i.e. 10x3s, 6x5s, 3x10s, 2x15s, 1x30s). These tests show that exposures equal to or longer than 3 seconds have minimal shutter timing effect ($<$ 1\%) and therefore we did not apply the shutter correction to any frame.

Several nights of data were reduced with both dome and sky flat fields, and the results of the standard star photometry using both flat fields were compared. Data reduced using dome flat fields had a smaller scatter in the fit to the standard stars than those reduced with the twilight-sky flat fields and therefore all the data were reduced using only dome flat fields.

Aperture photometry was performed on the standard star and program star observations after the bias subtraction, linearity and flat field correction. Different aperture sizes were tested using the standard stars to derive a photometric solution; a $6\farcs5$ radius (15-pixel) yields the smallest residual in the fit; therefore a $6\farcs5$ radius was used for all photometric data reduction. Most of the binaries have a separation of 5$\arcsec$-15$\arcsec$. In order to minimize the effect of light from the primary extending into the sky annulus of the secondary, an inner radius for the sky annulus was set to be $26\farcs1$ (60-pixel) and the width to be $2\farcs2$ (5-pixel). Few pairs have a separation of about 20$\arcsec$-30$\arcsec$; the inner radius of sky annulus was set to be $13\farcs0$ for those pairs.

An aperture correction was applied to all program stars using the \emph{apcor} task in IRAF. For each frame, about 10 bright but unsaturated and uncrowded stars were selected and used as a template for the aperture correction. This procedure was important for this work, particularly to derive reliable photometry for the secondaries, since the secondaries are much fainter than the primaries and the angular separations of the binaries are small. We first tested the aperture correction procedure on the standard star field and found that the difference with and without aperture correction on the standard star field is almost negligible. Therefore we believe that the aperture correction does not introduce any errors into the photometry of the program stars. For about 75\% of the pairs, photometric measurements of target stars were obtained using a $2\farcs6$  (6-pixel) radius aperture, which was then corrected to a $6\farcs5$  radius aperture, the aperture size used for the standard star photometry. For the remaining one quarter of the pairs (those separations less than 8$\arcsec$), a $1\farcs7$ (4-pixel) radius aperture size was used to minimize contamination from the bright primaries.  

The measured $BVRI$ instrumental magnitude of all standard stars observed during each night were used to derive a photometric solution for each night with a photometric zero point, a first-order extinction term, and a color term. On every night there were insufficient statistics to detect a second-order extinction coefficient; this term was held constant at zero. We also followed the suggestions of \cite{Harris1981} to solve for a new photometry solution over multiple nights: we averaged the previously derived color coefficients over each run and fed them back into the IRAF utility to fit for a new photometric zero point and extinction coefficient. The photometry derived from this new solution has negligible difference ( $<$ 0.015 mag in R band and $<$ 0.01 mag in BVI band) from that derived from the previous solution; we therefore keep the previous one. The photometric solutions for each night are shown in Table \ref{table:photsol}. Finally, the photometric solutions derived from the standard stars on each night was applied to all candidates.

\section{Results}\label{sec:results}

\subsection{$BVRI$ photometry and errors}\label{sec:phot}
Photometric measurements were obtained for 76 of the 80 candidate pairs. We report the $BVRI$ photometry of observed primaries in Table \ref{table:phot} and secondaries in Table \ref{table:phot2}. In Table \ref{table:phot} (Table \ref{table:phot2}), Column 1 gives the name of primary (or secondary) in the rNLTT catalog or LSPM catalog. Columns 2-9 present the $BVRI$ photometry for primaries (secondaries) measured in this program and their errors (more details are provided below in this section). Column 10-12 list the $JHK$ photometry for primaries (secondaries) from 2MASS \citep{Skrutskie2006}. Column 13 gives the number of measurements of each primary (secondary), while column 14 contains side notes. Multiple (i.e. two or more) observations were obtained of 75 targets among the 154 stars. When only one measurement of a target was obtained, the value from that measurement is reported. Photometry of stars with multiple measurements is determined by taking all average of the photometric measurements for that star.

We were unable to derive reliable photometry for seven primaries since they are too bright for our instrumental setup and were saturated in one or several bands (usually i-band) even with the shortest exposure time. For one pair PM I20072-3519E \& PM I20072-3519W, PM I20072-3519E was originally assigned to be the primary since its $V$ magnitude is brighter than PM I20072-3519W in LSPM catalog. However, our photometry in this work shows that PM I20072-3519W is actually brighter and bluer, so we take PM I20072-3519W as primary in Table \ref{table:phot} and PM I20072-3519E as secondary in Table \ref{table:phot2}. Also, the photometry from our measurements shows that eight pairs with primaries having $(V-J) > 2.5$ and they are very likely to be M+M dwarf pairs; also, seven pairs with secondaries having $(V-J) < 2.8$ and they are probably F/G/K+K pairs; we listed these pairs in Table \ref{table:sample} (see footnotes $k$ and $l$).

There are two types of errors considered in this program, systematic and statistical. There are several sources of systematic error in the measurements: cosmic rays or bad pixels may contaminate the measurements; the non-linearity issue mentioned previously may still have some residual effect even after the correction; the variation of the atmospheric conditions could also affect the fit to the standard star photometry and therefore increase the photometric errors. In this work, we take the standard deviation of the residual in the fit to the standard stars as the systematic error, since the S/N from those standard star measurements are adequately large.  This scatter from the standard star fitting is almost the same from night to night; we therefore use the average standard deviation of the residual from the fit to represent the systematic error in this program. They are 0.023, 0.027, 0.024, 0.028 mag in $V$, $B-V$, $V-R$, and $V-I$, respectively. However, most Landolt standards are bluer than the secondaries in our sample, so the systematic errors on the secondaries are potentially larger than the scatter from the standard star fitting.

\begin{figure}[htbp]
  \epsscale{1}
  \plotone{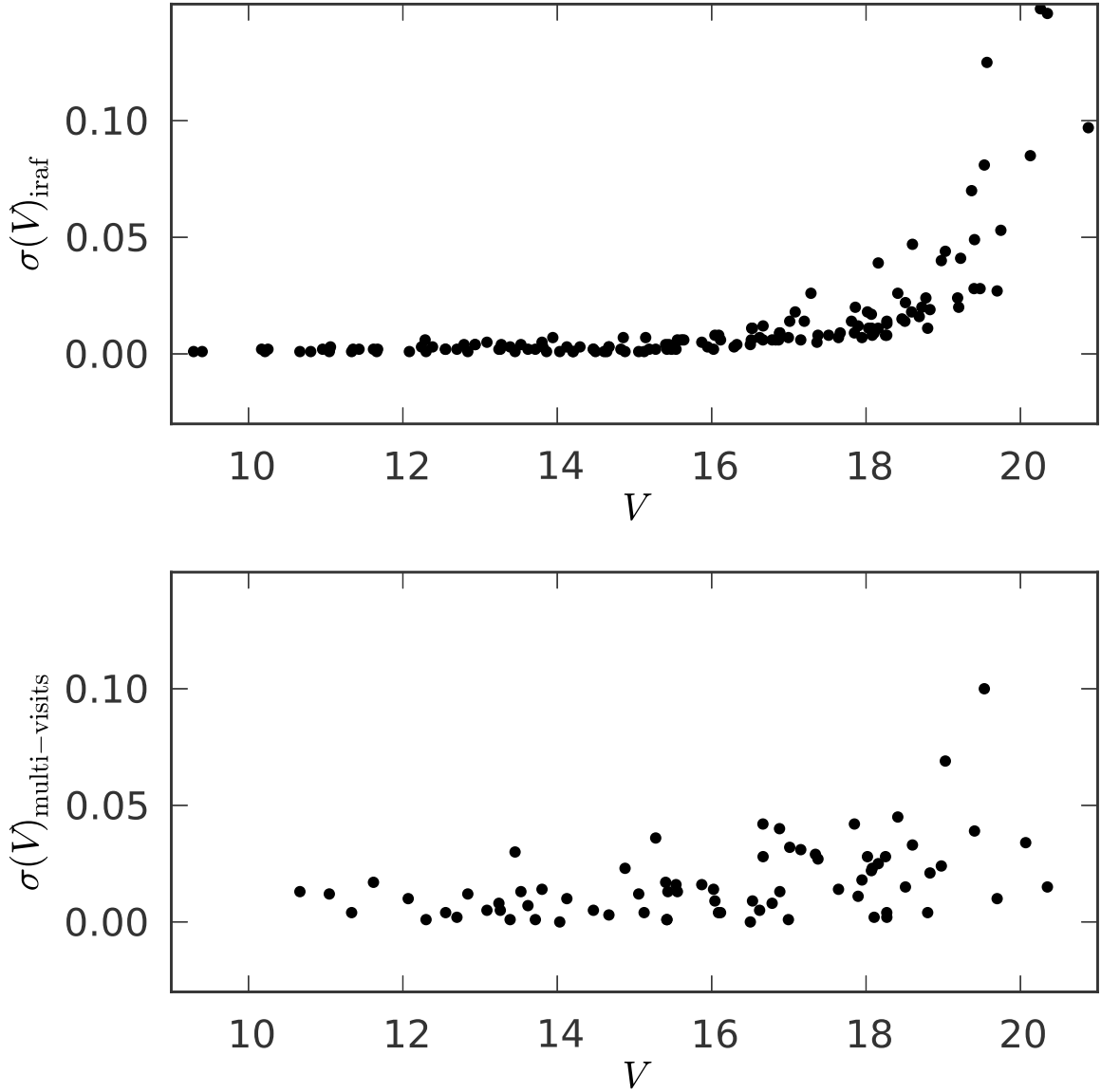}
  \caption{Top panel: the statistical errors of $V$ photometry as a function of $V$ magnitude. Photometric errors increase for fainter stars with lower S/N. Bottom panel: Standard deviations of the photometry for the 75 stars having multiple measurements as a function of $V$ magnitude. The standard deviations have a similar trend as the errors from the top panel but they are larger for brighter stars since the standard deviations are dominated by the systematic errors for $V< 16$ mag. }
  \label{fig:error}
\end{figure}

IRAF computes errors for each photometric measurement made using the \emph{phot} package. These errors are based on the number of electrons recorded by the detector, and decrease as the S/N of the measurements increases. We therefore take the error reported by IRAF as the statistical error. Since the sample has $9 < V < 21$ mag, the statistical error, $\sigma(V)$, also varies from 0.001 mag ($V \sim 10$ mag) to 0.25 mag ($V \sim 20$ mag) based on magnitude. The top panel of Figure \ref{fig:error} shows the statistical error of $V$ band photometry as a function of $V$ magnitude; we note the expected trend of increased error for the fainter stars. The secondaries in our sample are generally faint, red, M dwarfs; they are therefore quite faint in the B filter and the statistical error of $B-V$ can be as large as 1 magnitude for some of the faint secondaries. 

We tested the appropriateness of the estimated errors by comparing photometric measurements of the stars that have more than one measurement. The bottom panel of Figure \ref{fig:error} plots the standard deviation of the multiple measurements for 75 stars as a function of the measured $V$ magnitude. This standard deviation shows a similar trend as the error derived from the IRAF measurements shown in the top panel of Figure \ref{fig:error}. For $V <$ 16, where the statistical errors are negligible, the standard deviation of multiple measurements is about 0.01-0.02 mag, which matches the systematic errors $\sigma(V)$ $\sim$ 0.023. For $V>$ 18, the standard deviation of the multiple measurements gets larger since the errors are dominated by the statistical errors for faint stars. This shows that the error from IRAF is a good estimate of the statistical error.

The final error for a given candidate is reported in Table \ref{table:phot} or \ref{table:phot2} as the statistical error and the systematic error added in quadrature.

\subsection{Comparison with the Previously Published Photometry and an Improved RPM Diagram}\label{sec:comp}

\begin{figure}[htbp]
  \epsscale{1}
  \plotone{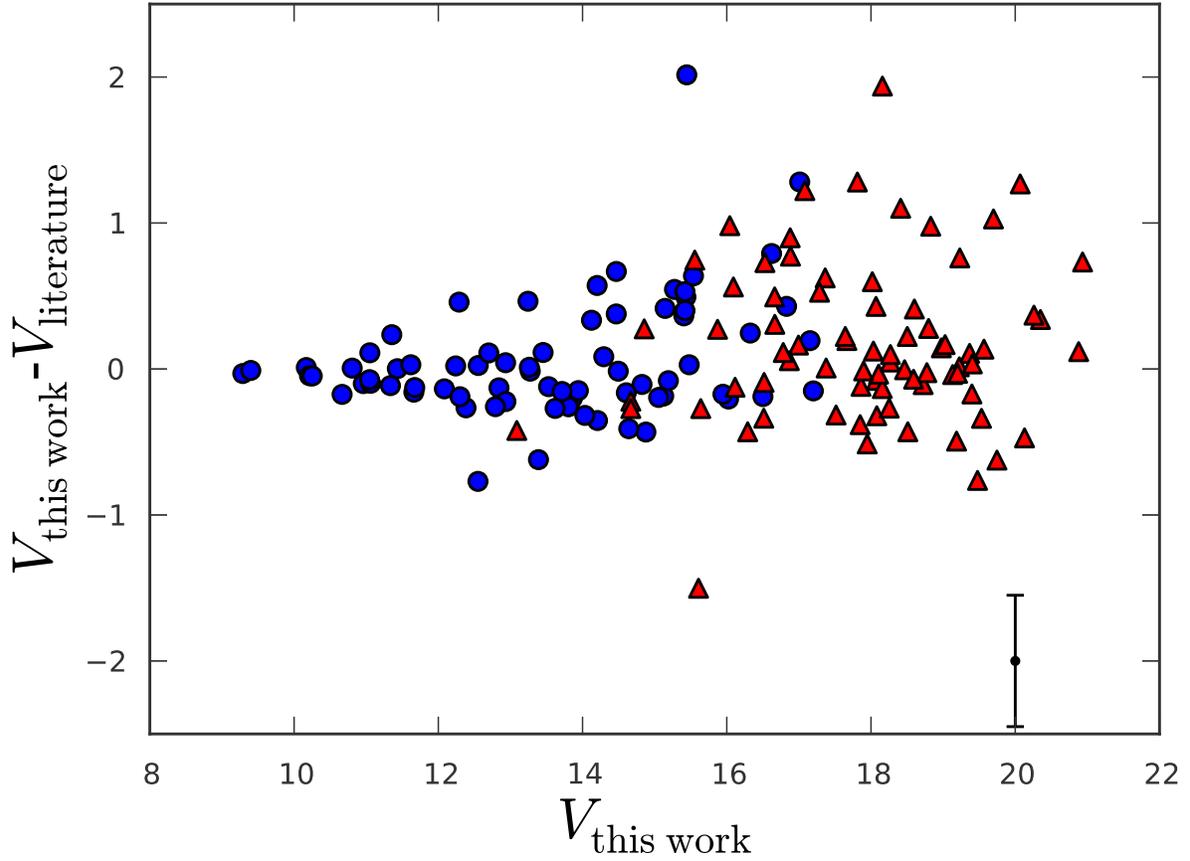}
  \caption{A comparison of the $V$ magnitude of the observed candidates to that of previously published photometry.  Symbols are the same as in Figure \ref{fig:rpm_old}. Primaries that have $V <$ 12 mag have a standard deviation of the difference of 0.098 mag, consistent with the Tycho-2 photometric errors of 0.10 mag \citep{Hog2000}. The difference for the total sample has a scatter of 0.52 mag in rms. This is also consistent with the rNLTT catalog and LSPM catalog photometric errors of about 0.45 mag in composite $V$ photometry, which is shown in the lower right corner.}
  \label{fig:compare_mag}
\end{figure}

Figure \ref{fig:compare_mag} compares the $V$ magnitude of the observed candidates to the photometry presented in the previously published catalogs. In both the rNLTT and LSPM catalog, the photometry is taken from the Tycho-2 catalog for most stars with $V <$ 12 mag; the remaining photometry is taken from the USNO-A for rNLTT and USNO-B1.0 for LSPM catalog. Seventeen primaries have $V <$ 12 mag. The standard deviation of the difference between this work and the previously published photometry for these 17 stars is 0.098 mag, which is consistent to the Tycho-2 photometric errors of 0.10 mag \citep{Hog2000}. The difference between the photometry presented here and the previously published photometry has a scatter of 0.52 mag for the whole sample. This is not surprising since the composite $V$ photometry from rNLTT catalog and LSPM catalog has photometric errors of about 0.45 mag \citep[USNO-A and USNO-B catalogs has photometry errors of 0.3 mag in each band,][]{monet2003}. 

\begin{figure}[htbp]
  \epsscale{1}
  \plotone{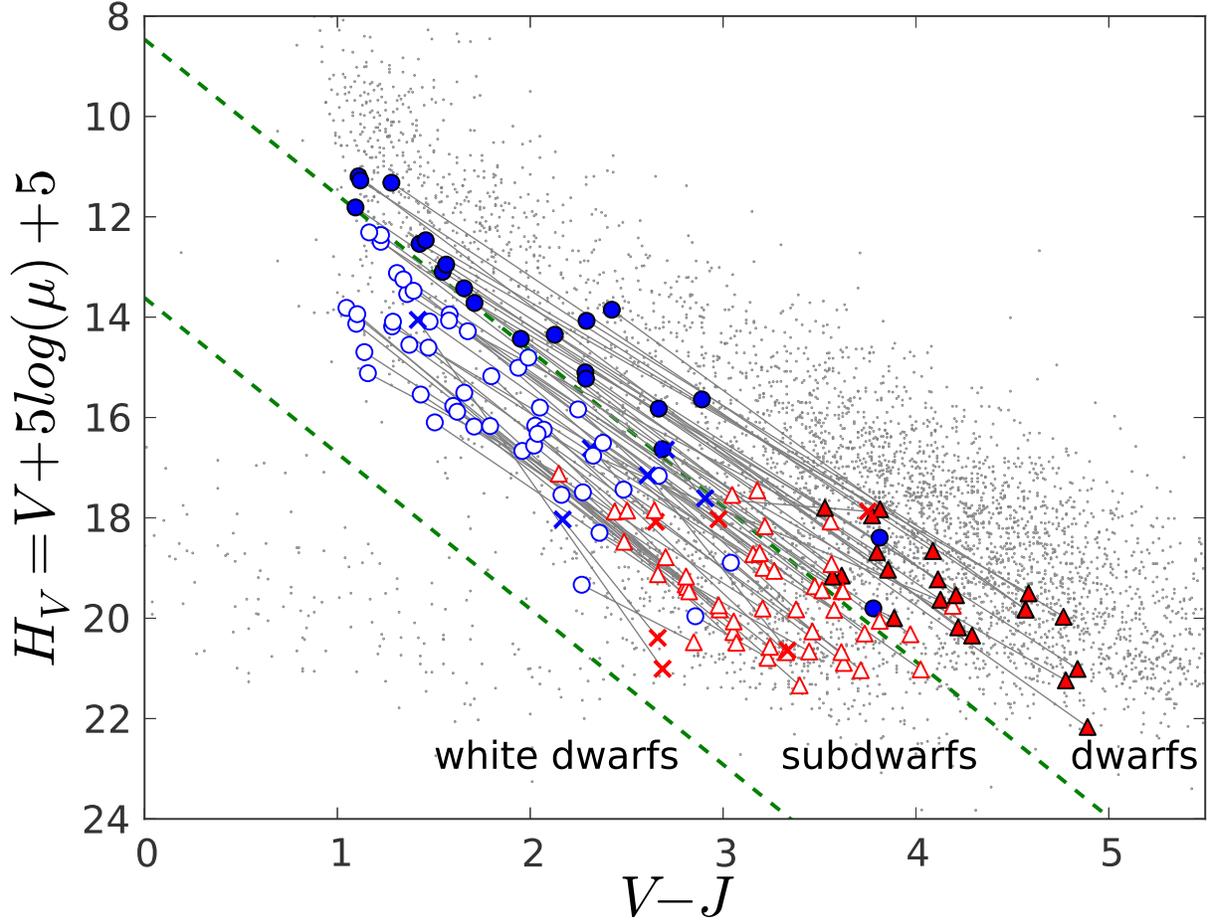}
  \caption{RPM diagram of the 74 pairs having improved $V$ photometry for both primaries (blue) and secondaries (red). A line connects the primary and secondary of each pair and the slope of each connected line is calculated. Six pairs with slopes $m < 1.5$ or $m > 4$ are unlikely to be pairs composed of two stars that have similar metallicities. These six pairs are plotted as crosses. Among the remaining 68 pairs after removal of the potential non-binaries, twenty-one pairs that have primaries above the discriminator line $\eta=0$ (dwarfs) are plotted as filled circles (primaries) and triangles (secondaries); forty-seven pairs that have primaries below the discriminator line $\eta=0$ (subdwarfs), are plotted as open circles (primaries) and triangles (secondaries). Overplotted grey dots are stars from the LSPM-North catalog.}
  \label{fig:rpm_new}
\end{figure}

\begin{figure}[htbp]
  \epsscale{1}
  \plotone{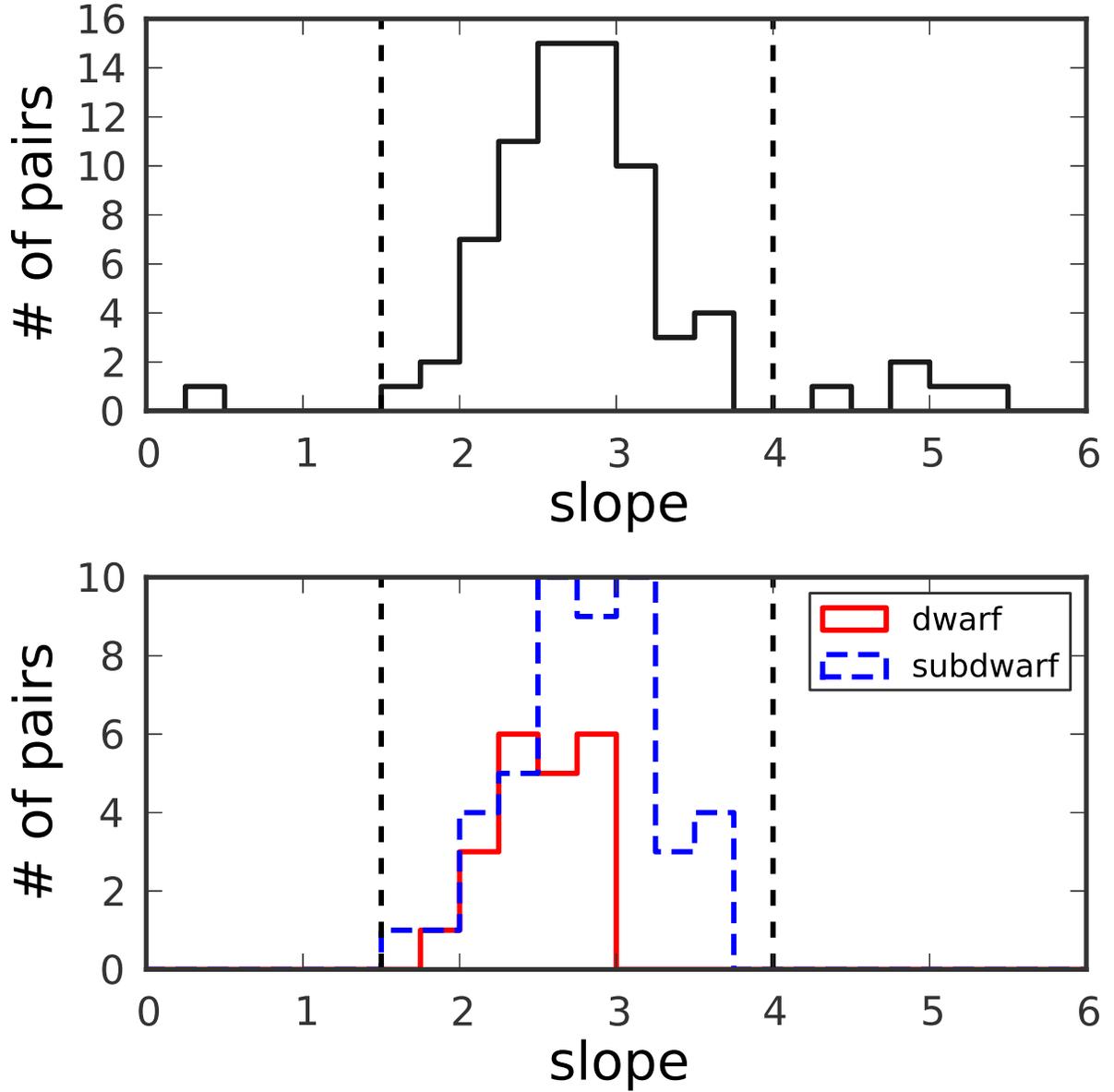}
  \caption{Top Panel: A histogram of the slopes of the connected lines in Figure \ref{fig:rpm_new}. Six pairs with slopes $m < 1.5$ or $m > 4$ are unlikely to be pairs composed of two stars that have similar metallicities. Bottom Panel: Histograms of the slopes $m$ for two groups. Red curve represents the dwarf pair candidates (filled symbols in Figure \ref{fig:rpm_new}) and blue one indicates the subdwarf pairs candidates (open symbols in Figure \ref{fig:rpm_new}). The dwarf candidates in general have a smaller slope than the subdwarf candidates. }
  \label{fig:slope_hist}
\end{figure}

An improved RPM diagram constructed from the photometry presented in this work is given in Figure \ref{fig:rpm_new}. Again, the $J$ magnitude is from the 2MASS catalog. Seventy-four pairs have new measured photometry in $V$ for both primaries and secondaries and are plotted here. As in Figure \ref{fig:rpm_old}, a line connects the primary and secondary of each pair. Since we have selected common proper motion pairs whose two members are likely wide binaries and share the same metallicity, one expects that the line connecting their positions on the RPM diagram should be approximately parallel to the corresponding MS or SD track (i.e. slope $m\approx3.1$ as in Equation \ref{eq:rpmeq}). Some ``pairs'' do not follow this ``parallel rule''; these could have a WD or evolved companion \citep[see e.g.][]{Chaname2004}, or they could be co-moving pairs that are not actual wide binaries but just chance alignments on the sky. A histogram of the slopes of the connected lines, $m$, is plotted in the top panel of Figure \ref{fig:slope_hist}. Six pairs with slope $m < 1.5$ or $m > 4$ are unlikely to be binaries (and are marked as crosses in Figure \ref{fig:rpm_new}) and we do not consider them further in this work. We also listed these six pairs in Table \ref{table:sample} (see footnote $m$). Please note these criteria are arbitrarily selected for this sample since the majority of the slopes fall between $1.5 < m < 4$. The remaining 68 pairs are identified as most likely to be ``true" binaries, and are thus most likely to have the same metallicities for each pair. These potential F/G/K+M pairs will be the targets for a future spectroscopic study in which we will calibrate the metallicity of metal-poor M dwarfs (i.e. the M subdwarfs). In this future work, the assumption of a true binary pair will be further tested by measuring the radial velocities of both stars in each binary. 

As mentioned earlier, considering the large errors of the previously published photometry, we made a loose cut to include all potential subdwarf candidates, since the ultimate goal of this work was to provide a metallicity calibration for metal-poor M subdwarfs. As a result, about 30\% of the pairs still lie above the MS and SD discriminator line  $\eta=0$  (see Equation \ref{eq:rpmeq}) in the improved RPM diagram, which implies that they are likely to be dwarfs rather than subdwarfs. We will include these dwarfs in the spectroscopic study, since those pairs will provide a subset of metal-rich stars which will be useful to connect our calibration at the low-metallicity end to the high-metallicity calibrations that have already been determined for M dwarfs \citep[e.g.][]{Newton2014}.

Based on the MS and SD discriminator line $\eta=0$, we separate the sample into two groups, dwarf candidates and subdwarf candidates, using the improved RPM diagram; the candidate dwarf pairs with primaries lying above the discriminator line  $\eta=0$ are plotted as filled circles (primaries) and triangles (secondaries) in Figure \ref{fig:rpm_new}, while the candidate subdwarf pairs with primaries lying below the discriminator line $\eta=0$, are plotted as open circles (primaries) and triangles (secondaries). We also plot histograms of the slopes $m$ for the two groups in Figure \ref{fig:slope_hist}. The dwarf pairs in general have a smaller slope than the subdwarf pairs. This agrees with the fact that the color-magnitude relationship is expected to be shallower at high metallicity.

\subsection{Color-color Diagram and Color Dependence on Metallicity }\label{sec:color} 

\begin{figure}[htbp]
  \epsscale{1}
  \plotone{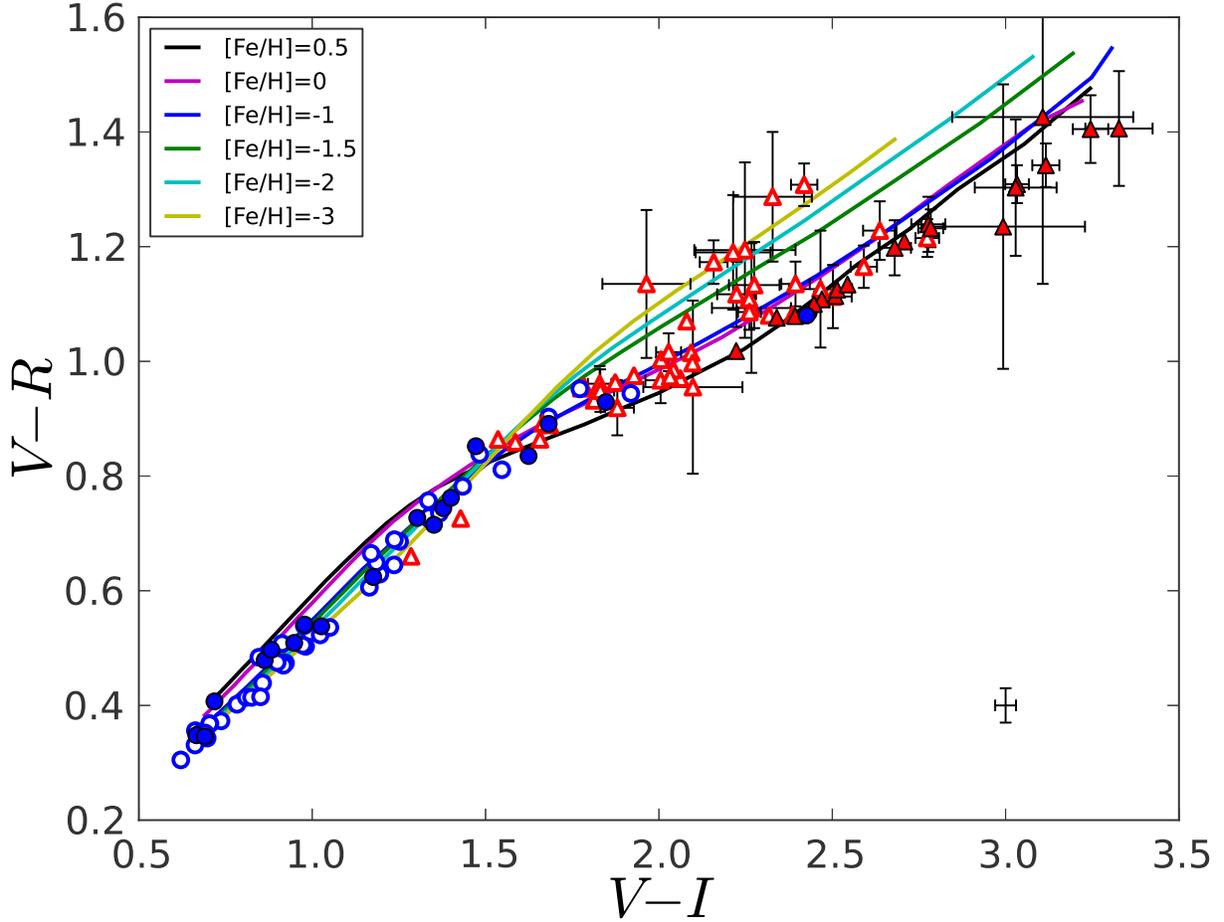}
  \caption{VRI color-color diagram for the 68 pairs. Symbols are the same as in Figure \ref{fig:rpm_new}. Typical errors of 0.03 mag are shown in the lower right corner. Colors with errors larger than 0.03 mag are shown with associated error bars. Open triangles generally lie above the filled triangles for the secondaries, which shows that the M subdwarf candidates have a redder $V-R$ color for a given $V-I$ color.  Overplotted are synthetic colors generated from the PHOENIX/BT-Settl model \citep{Allard2011} for $\log g = 5$, $T_{eff}$ = 2800K-6000K and various metallicities from $\rm{[Fe/H]}=0.5$ to $\rm{[Fe/H]}=-3$, as indicated by the legend. The models corroborate the observations and predict that the M subdwarf candidates are likely to indeed be more metal-poor than the M dwarf candidates.  }
  \label{fig:VRI}
\end{figure}

We have constructed a color-color diagram in the $V-R$ vs. $V-I$ plane for the 68 potential ``true'' pairs in Figure \ref{fig:VRI}. Symbols are the same as in Figure \ref{fig:rpm_new}. In Figure \ref{fig:VRI}, the subdwarf candidates mainly lie above the dwarf candidates on this diagram; i.e., the subdwarfs generally have a redder $V-R$ color at a given $V-I$ color. This appears to be analogous to the apparent  correlation between metallicity and $g-r$ color found for M dwarfs and M subdwarfs by \cite{Lepine2008b} and \cite{Bochanski2013}. The color dependence of M dwarfs and subdwarfs on metallicity is also predicted by modern atmospheric models. Overplotted on Figure \ref{fig:VRI} are synthetic colors generated from the PHOENIX/BT-Settl\footnote{http://perso.ens-lyon.fr/france.allard/} model \citep{Allard2011} for $\log g = 5$, $T_{eff}$ = 2800K-6000K and various metallicities from $\rm{[Fe/H]}=0.5$ to $\rm{[Fe/H]}=-3$. These models corroborate the observations and predict that the M subdwarfs generally have redder $V-R$ colors than the M dwarfs. 

In future work, we will obtain more precise metallicities of the primaries in the sample using high-resolution spectra (Marshall et al. 2014, in preparation). The metallicities, together with the precisely measured photometry presented in this work, will place tighter constraints on the models. Moreover, the accurate metallicities derived from the primaries could also help develop a better relationship between the metallicities and the colors of the M dwarfs and subdwarfs, which could be used to identify more metal-poor, low mass stars in the vicinity of the Sun.

\section{Summary and Future Work}\label{sec:concl}
We present the improved $BVRI$ photometry for a sample of high common proper motion F/G/K+M subdwarf  pairs selected from a RPM diagram. These measurements are generally precise to 0.03 mag, representing a significant improvement on photometry currently existing for these stars.  We have presented a revised RPM diagram for these candidate pairs using our improved photometry. A $V-R$ vs $V-I$ plot is also constructed using the improved photometry. There is a photometric dependence on the metallicity for the M dwarf and subdwarf candidates, where the subdwarf candidates have generally redder $V-R$ colors at a given $V-I$ color, as predicted by modern atmospheric models. 

In a future paper, we will present metallicities for the primaries of our sample derived from high-resolution spectra along with line indices for the secondaries measured in moderate-resolution spectra. Together with the improved photometry presented here, we will develop an empirical calibration of M subdwarf metallicities. Moreover, the accurate metallicity derived from primaries could also help develop a better relationship between the metallicities and the colors of the M dwarfs and subdwarfs. A photometric metallicity method like this could be used to identify more low mass subdwarfs in the vicinity of the Sun in future photometric surveys with high photometric precision, such as the Large Synoptic Survey Telescope \citep{Ivezic2008,Ivezic2012}.

\acknowledgements
T. Li wishes to thank D. L. DePoy and Dan Q. Nagasawa for the fruitful conversations and discussions.

%
%
%
%
%
%
%
%
%
%
%
%
%
%
%
%
%
%
%
%
%
%
%
%
%
%
%
%
%
%
%
%
%
%
%
%
%
%
%
%
%
%
%
%
%
%
%
%
%
%

\clearpage
\begin{deluxetable}{lcccccc|lcccccc|cc}
\setlength{\tabcolsep}{0.035in}
\rotate
\tabletypesize{\scriptsize}
\tablecolumns{16}
\tablewidth{0pc}
\tablecaption{
F/G/K+M  Common Proper Motion Sample
\label{table:sample}
}
\tablehead{
\colhead{Primary} & \colhead{} & \colhead{} & \colhead{} & \colhead{} & \colhead{} & \colhead{} & \colhead{Secondary} & \colhead{} & \colhead{} & \colhead{} & \colhead{} &  \colhead{}  &  \colhead{}\\
 \colhead{ID\tablenotemark{a}} & \colhead{RA} & \colhead{Dec} & \colhead{pmRA\tablenotemark{b}} & \colhead{pmDEC\tablenotemark{b}} & \colhead{V\tablenotemark{c}} & \colhead{V-J\tablenotemark{c}} & \colhead{ID\tablenotemark{a}} &  \colhead{RA} & \colhead{Dec} & \colhead{pmRA\tablenotemark{b}} & \colhead{pmDEC\tablenotemark{b}} & \colhead{V\tablenotemark{c}} & \colhead{V-J\tablenotemark{c}} & \colhead{Sep\tablenotemark{d}} & \colhead{Notes}
}
\startdata
PM I00025-4644 & 00:02:35.66 & -46:44:52.0 & 0.150 & -0.049 & 16.69 & 2.47 & PM I00026-4644 & 00:02:36.21 & -46:44:57.9 & 0.150 & -0.049 & 20.01 & 3.06 & 8.1 &  \\
PM I00329+1805 & 00:32:55.80 & +18:05:52.9 & 0.095 & -0.065 & 16.08 & 2.44 & PM I00329+1805-2 & 00:32:56.85 & +18:05:56.5 & 0.095 & -0.065 & 20.20 & 4.04 & 15.4 & \tablenotemark{e,k} \\
PM I00422+0731E & 00:42:15.23 & +07:31:18.7 & 0.169 & -0.068 & 15.31 & 2.14 & PM I00422+0731W & 00:42:14.35 & +07:31:19.9 & 0.169 & -0.068 & 18.83 & 2.91 & 13.1 &  \\
PM I00592+0705N & 00:59:17.81 & +07:05:56.4 & 0.116 & -0.003 & 14.06 & 1.48 & PM I00592+0705S & 00:59:17.38 & +07:05:47.0 & 0.103 & -0.017 & 19.26 & 3.51 & 11.4 &  \\
NLTT 3847 & 01:09:28.97 & -05:07:25.3 & 0.743 & 0.044 & 13.43 & 1.78 & NLTT 3849 & 01:09:29.34 & -05:07:30.6 & 0.743 & 0.044 & 15.91 & 4.19 & 8.0 & \tablenotemark{k} \\
PM I01227+1409 & 01:22:43.29 & +14:09:34.5 & -0.078 & -0.140 & 10.16 & 1.10 & PM I01226+1409E & 01:22:41.13 & +14:09:28.8 & -0.077 & -0.135 & 19.21 & 4.20 & 31.9 &  \\
PM I01266-4842W & 01:26:37.33 & -48:42:51.0 & 0.205 & -0.050 & 13.63 & 2.09 & PM I01266-4842E & 01:26:38.28 & -48:42:54.9 & 0.205 & -0.050 & 18.83 & 4.40 & 10.2 & \tablenotemark{k} \\
NLTT 4817 & 01:26:55.17 & +12:00:25.9 & -0.014 & -0.359 & 11.12 & 0.89 & NLTT 4814 & 01:26:54.13 & +12:00:06.8 & -0.011 & -0.364 & 16.61 & 2.88 & 24.4 &  \\
PM I01352+0538N & 01:35:14.71 & +05:38:24.7 & 0.285 & -0.035 & 10.80 & 1.54 & PM I01352+0538S & 01:35:14.24 & +05:38:12.0 & 0.267 & -0.026 & 15.86 & 2.89 & 14.4 &  \\
PM I01430-4959W & 01:43:00.68 & -49:59:26.8 & 0.021 & -0.181 & 14.09 & 1.22 & PM I01430-4959E & 01:43:01.27 & -49:59:22.1 & 0.021 & -0.181 & 18.15 & 2.89 & 7.4 &  \\
PM I02012+0218 & 02:01:15.09 & +02:18:25.8 & 0.209 & -0.024 & 17.44 & 2.25 & PM I02012+0217 & 02:01:17.12 & +02:17:29.7 & 0.209 & -0.024 & 18.87 & 3.28 & 63.8 & \tablenotemark{h} \\
PM I02225+1531S & 02:22:34.06 & +15:31:09.9 & 0.188 & -0.141 & 12.67 & 1.66 & PM I02225+1531N & 02:22:33.00 & +15:31:47.8 & 0.188 & -0.141 & 18.33 & 3.46 & 40.8 & \tablenotemark{h} \\
PM I02267-4214 & 02:26:47.96 & -42:14:58.9 & -0.022 & -0.160 & 13.85 & 1.54 & PM I02267-4215 & 02:26:46.85 & -42:15:06.8 & -0.022 & -0.160 & 19.25 & 3.90 & 14.7 & \tablenotemark{h} \\
NLTT 8753 & 02:42:05.13 & -24:45:16.3 & -0.058 & -0.419 & 16.40 & 2.42 & NLTT 8759 & 02:42:14.98 & -24:44:18.0 & -0.056 & -0.418 & 17.47 & 3.02 & 146.3 & \tablenotemark{k}  \\
PM I02548+2057W & 02:54:49.43 & +20:57:34.8 & 0.040 & -0.115 & 16.54 & 2.39 & PM I02548+2057E & 02:54:50.04 & +20:57:32.1 & 0.036 & -0.122 & 19.86 & 3.29 & 8.9 & \tablenotemark{h} \\
PM I02569-5831N & 02:56:55.71 & -58:31:24.3 & 0.152 & -0.098 & 14.07 & 2.54 & PM I02569-5831S & 02:56:56.81 & -58:31:36.6 & 0.152 & -0.098 & 20.76 & 4.77 & 15.0 &  \\
PM I03150+0102 & 03:15:04.76 & +01:02:15.2 & 0.363 & 0.117 & 10.26 & 1.36 & PM I03150+0103 & 03:15:00.92 & +01:03:08.3 & 0.353 & 0.128 & 14.89 & 3.27 & 78.3 &  \\
PM I03256-3333E & 03:25:41.79 & -33:33:34.6 & 0.137 & -0.182 & 14.93 & 2.42 & PM I03256-3333Wn & 03:25:40.93 & -33:33:25.3 & 0.137 & -0.182 & 16.72 & 3.07 & 14.2 & \tablenotemark{m} \\
NLTT 12296 & 03:59:04.27 & -06:56:03.2 & 0.024 & -0.253 & 15.05 & 2.42 & NLTT 12294 & 03:59:02.49 & -06:56:33.8 & 0.023 & -0.258 & 18.28 & 3.01 & 40.5 &  \\
PM I04072+1526N & 04:07:16.36 & +15:26:42.8 & 0.109 & -0.122 & 10.30 & 1.33 & PM I04072+1526S & 04:07:15.45 & +15:26:20.2 & 0.103 & -0.120 & 17.85 & 3.59 & 26.2 &  \\
PM I04099+0942E & 04:09:54.30 & +09:42:58.8 & 0.078 & -0.273 & 14.21 & 1.94 & PM I04099+0942W & 04:09:54.07 & +09:42:56.4 & 0.078 & -0.273 & 17.11 & 3.94 & 4.2 & \tablenotemark{l} \\
PM I04254-4601 & 04:25:28.74 & -46:01:23.9 & 0.100 & -0.124 & 14.51 & 1.67 & PM I04255-4601 & 04:25:30.84 & -46:01:22.7 & 0.100 & -0.124 & 20.24 & 3.83 & 21.9 &  \\
PM I04325-5657N & 04:32:32.44 & -56:57:04.3 & 0.124 & -0.133 & 12.22 & 1.34 & PM I04325-5657S & 04:32:31.79 & -56:57:14.6 & 0.124 & -0.133 & 16.81 & 3.16 & 11.6 &  \\
PM I04327+0820 & 04:32:45.59 & +08:20:05.5 & 0.053 & -0.085 & 14.09 & 1.73 & PM I04327+0820-2 & 04:32:46.10 & +08:20:14.6 & 0.053 & -0.085 & 19.89 & 3.09 & 11.9 & \tablenotemark{e} \\
PM I04332+0013 & 04:33:17.84 & +00:13:59.8 & -0.060 & -0.152 & 11.43 & 1.22 & PM I04333+0014 & 04:33:18.67 & +00:14:14.0 & -0.064 & -0.140 & 16.85 & 3.51 & 18.9 &  \\
PM I04477-3044W & 04:47:42.65 & -30:44:03.2 & 0.157 & -0.142 & 11.59 & 1.31 & PM I04477-3044E & 04:47:44.22 & -30:44:02.5 & 0.143 & -0.143 & 18.80 & 3.75 & 20.3 & \tablenotemark{i} \\
NLTT 14407 & 05:02:20.19 & -19:32:04.4 & 0.527 & -0.419 & 11.82 & 2.20 & NLTT 14408 & 05:02:20.95 & -19:32:01.9 & 0.527 & -0.419 & 14.58 & 2.90 & 11.0 &  \\
PM I05137+0647W & 05:13:46.03 & +06:47:01.0 & 0.134 & -0.109 & 11.83 & 0.94 & PM I05137+0647E & 05:13:47.17 & +06:47:08.6 & 0.134 & -0.109 & 19.17 & 4.01 & 18.6 &  \\
PM I05195+0903E & 05:19:34.77 & +09:03:46.3 & 0.079 & -0.234 & 14.56 & 2.15 & PM I05195+0903W & 05:19:34.09 & +09:03:36.8 & 0.079 & -0.234 & 17.98 & 3.10 & 13.9 &  \\
PM I05484-3617Nn & 05:48:28.64 & -36:17:06.7 & 0.090 & -0.156 & 14.77 & 1.78 & PM I05484-3617S & 05:48:29.65 & -36:17:19.5 & 0.090 & -0.156 & 18.47 & 2.98 & 17.6 &  \\
PM I06032+1921N & 06:03:14.87 & +19:21:38.6 & 0.667 & -0.623 & 9.32 & 1.32 & PM I06032+1921S & 06:03:14.51 & +19:21:34.0 & 0.659 & -0.609 & 13.51 & 2.92 & 6.9 &  \tablenotemark{l} \\
PM I06050+0723S & 06:05:03.52 & +07:23:30.5 & -0.199 & -0.098 & 14.94 & 2.17 & PM I06050+0723N & 06:05:03.39 & +07:23:39.1 & -0.199 & -0.098 & 16.83 & 2.99 & 8.8 & \tablenotemark{k}  \\
PM I06394-3030E & 06:39:24.52 & -30:30:50.9 & 0.040 & 0.183 & 14.73 & 2.16 & PM I06394-3030W & 06:39:24.03 & -30:30:55.3 & 0.040 & 0.183 & 16.17 & 2.48 & 7.0 & \tablenotemark{m} \\
PM I06436+0851 & 06:43:36.55 & +08:51:44.4 & 0.069 & -0.055 & 13.79 & 2.09 & PM I06436+0851-2 & 06:43:35.88 & +08:51:40.8 & 0.084 & -0.065 & 18.67 & 3.54 & 10.5 & \tablenotemark{e} \\
PM I08152-6337 & 08:15:17.60 & -63:37:18.4 & -0.152 & 0.240 & 12.22 & 2.26 & PM I08153-6337 & 08:15:18.03 & -63:37:04.5 & -0.152 & 0.240 & 16.74 & 3.50 & 14.2 &  \\
PM I08239-7549W & 08:23:54.94 & -75:49:34.3 & -0.045 & 0.202 & 11.06 & 1.53 & PM I08239-7549E & 08:23:58.54 & -75:49:32.4 & -0.030 & 0.216 & 16.53 & 3.30 & 13.3 &  \\
PM I08386-3856 & 08:38:36.72 & -38:56:55.7 & 0.101 & -0.126 & 12.65 & 1.92 & PM I08386-3857 & 08:38:37.73 & -38:57:14.4 & 0.101 & -0.126 & 17.42 & 3.57 & 22.0 &  \\
PM I09502+0509E & 09:50:13.89 & +05:09:02.4 & 0.210 & -0.220 & 11.80 & 1.61 & PM I09502+0509W & 09:50:12.89 & +05:09:08.2 & 0.217 & -0.216 & 18.66 & 4.09 & 16.0 &  \\
PM I10105+1203W & 10:10:34.78 & +12:03:17.6 & -0.185 & -0.078 & 12.53 & 1.55 & PM I10105+1203E & 10:10:35.17 & +12:03:22.9 & -0.185 & -0.078 & 18.23 & 3.86 & 7.8 &  \\
PM I10520+1521N & 10:52:02.16 & +15:21:18.6 & 0.134 & -0.150 & 16.23 & 2.37 & PM I10520+1521S & 10:52:01.65 & +15:21:09.5 & 0.134 & -0.150 & 18.46 & 3.34 & 11.7 &  \\
PM I11110-4414 & 11:11:04.09 & -44:14:15.9 & -0.134 & -0.320 & 12.97 & 1.56 & PM I11109-4416 & 11:10:57.83 & -44:16:31.1 & -0.152 & -0.306 & 16.24 & 2.83 & 151.0 &  \tablenotemark{l}  \\
PM I11125-3512 & 11:12:30.07 & -35:12:35.0 & -0.115 & -0.113 & 15.31 & 2.20 & PM I11124-3512 & 11:12:28.92 & -35:12:36.0 & -0.115 & -0.113 & 18.52 & 3.30 & 14.1 &  \\
NLTT 27188 & 11:22:26.44 & -27:13:35.2 & -0.320 & 0.063 & 13.32 & 1.93 & NLTT 27182 & 11:22:23.60 & -27:13:45.0 & -0.302 & 0.066 & 14.94 & 2.42 & 39.1 &  \tablenotemark{l}  \\
PM I11263+2047Ee & 11:26:21.55 & +20:47:22.7 & 0.118 & -0.193 & 13.34 & 2.18 & PM I11263+2047Ew & 11:26:20.37 & +20:47:15.0 & 0.118 & -0.193 & 17.37 & 3.61 & 18.2 &  \\
PM I11330+1318N & 11:33:02.86 & +13:18:33.2 & -0.237 & -0.010 & 9.41 & 1.13 & PM I11330+1318S & 11:33:03.95 & +13:18:17.1 & -0.227 & 0.008 & 15.98 & 3.19 & 22.7 &  \\
PM I11392-4118N & 11:39:12.23 & -41:18:15.1 & -0.206 & 0.220 & 12.78 & 2.43 & PM I11392-4118S & 11:39:12.30 & -41:18:26.2 & -0.206 & 0.220 & 14.81 & 3.03 & 11.1 & \tablenotemark{k}  \\
PM I11584-4155E & 11:58:27.99 & -41:55:19.3 & -0.769 & -0.266 & 9.00 & 1.67 & PM I11584-4155W & 11:58:26.46 & -41:55:03.4 & -0.766 & -0.274 & 15.06 & 3.28 & 23.3 &  \\
PM I12170+0742E & 12:17:05.76 & +07:42:30.3 & 0.034 & -0.158 & 15.25 & 1.70 & PM I12170+0742W & 12:17:04.98 & +07:42:31.1 & 0.034 & -0.158 & 18.40 & 2.98 & 11.6 &  \tablenotemark{l}  \\
PM I12237+0625 & 12:23:43.48 & +06:25:10.3 & -0.173 & -0.021 & 14.06 & 2.20 & PM I12237+0624 & 12:23:44.04 & +06:24:48.4 & -0.173 & -0.021 & 18.52 & 3.89 & 23.4 &  \\
PM I12277+1334 & 12:27:43.78 & +13:34:16.2 & 0.082 & -0.248 & 16.12 & 2.34 & PM I12277+1336 & 12:27:46.62 & +13:36:37.0 & 0.077 & -0.269 & 18.29 & 2.80 & 146.7 & \tablenotemark{m} \\
PM I12283+1222S & 12:28:18.28 & +12:22:36.4 & -0.184 & -0.127 & 12.67 & 1.27 & PM I12283+1222N & 12:28:18.75 & +12:22:50.7 & -0.160 & -0.112 & 19.23 & 3.21 & 15.8 & \tablenotemark{j} \\
PM I12440+0625E & 12:44:02.57 & +06:25:46.9 & 0.065 & -0.135 & 13.16 & 1.27 & PM I12440+0625We & 12:44:00.58 & +06:25:48.3 & 0.065 & -0.135 & 19.68 & 3.55 & 29.7 &  \\
PM I12508+0757 & 12:50:48.80 & +07:57:56.7 & 0.047 & -0.163 & 10.84 & 1.27 & PM I12507+0758 & 12:50:47.08 & +07:58:08.0 & 0.027 & -0.151 & 18.14 & 3.89 & 28.0 &  \\
PM I13116+1106 & 13:11:41.81 & +11:06:24.8 & -0.102 & -0.122 & 12.89 & 1.06 & PM I13116+1105 & 13:11:32.30 & +11:05:40.7 & -0.106 & -0.114 & 17.83 & 2.80 & 146.8 &  \tablenotemark{l}  \\
NLTT 33282 & 13:13:09.08 & -07:42:15.2 & -0.163 & -0.196 & 15.04 & 2.15 & NLTT 33283 & 13:13:09.60 & -07:42:07.8 & -0.159 & -0.190 & 15.60 & 2.40 & 10.0 &  \tablenotemark{l}  \\
PM I13133-4153N & 13:13:20.50 & -41:53:14.0 & -0.145 & -0.044 & 13.65 & 2.07 & PM I13133-4153S & 13:13:21.15 & -41:53:29.2 & -0.145 & -0.044 & 18.22 & 3.52 & 16.8 &  \\
PM I13167+0810E & 13:16:47.28 & +08:10:27.4 & 0.189 & -0.146 & 11.16 & 1.66 & PM I13167+0810W & 13:16:45.12 & +08:10:21.5 & 0.186 & -0.130 & 18.47 & 4.08 & 32.7 &  \\
PM I13372-4244E & 13:37:14.25 & -42:44:54.8 & 0.129 & -0.104 & 17.35 & 2.51 & PM I13372-4244W & 13:37:13.54 & -42:44:54.4 & 0.129 & -0.104 & 19.43 & 3.31 & 7.8 &  \\
PM I14055+0244S & 14:05:30.97 & +02:44:23.4 & -0.055 & -0.131 & 15.45 & 2.04 & PM I14055+0244N & 14:05:31.65 & +02:44:35.1 & -0.055 & -0.131 & 20.60 & 4.10 & 15.5 &  \\
PM I14124+0517S & 14:12:28.75 & +05:17:28.5 & -0.053 & -0.172 & 13.29 & 1.39 & PM I14124+0517N & 14:12:28.04 & +05:17:40.1 & -0.053 & -0.172 & 19.57 & 3.78 & 15.7 &  \\
PM I14136-3634E & 14:13:41.61 & -36:34:39.2 & -0.030 & -0.166 & 15.27 & 2.12 & PM I14136-3634W & 14:13:40.94 & -36:34:43.6 & -0.030 & -0.166 & 17.91 & 2.68 & 9.1 &  \\
PM I14475+1134 & 14:47:35.80 & +11:34:13.7 & -0.049 & -0.172 & 13.05 & 1.67 & PM I14476+1134 & 14:47:36.16 & +11:34:36.3 & -0.049 & -0.172 & 20.37 & 3.25 & 23.1 & \tablenotemark{f,m} \\
PM I15413+1349N & 15:41:19.36 & +13:49:28.5 & -0.190 & -0.529 & 14.73 & 2.63 & PM I15413+1349S & 15:41:19.36 & +13:49:23.6 & -0.190 & -0.529 & 16.76 & 3.19 & 4.0 & \tablenotemark{k}  \\
PM I16008+0146E & 16:00:53.85 & +01:46:16.5 & -0.155 & -0.132 & 13.25 & 1.98 & PM I16008+0146W & 16:00:53.48 & +01:46:19.4 & -0.155 & -0.132 & 17.91 & 3.53 & 6.2 &  \\
PM I16519-4806N & 16:51:58.19 & -48:06:13.7 & -0.150 & -0.109 & 15.02 & 1.93 & PM I16519-4806S & 16:51:57.91 & -48:06:19.2 & -0.150 & -0.109 & 17.31 & 3.09 & 6.2 &  \\
PM I17135+1909 & 17:13:30.36 & +19:09:57.2 & -0.116 & -0.153 & 10.94 & 1.35 & PM I17134+1910 & 17:13:29.74 & +19:10:10.0 & -0.124 & -0.133 & 15.80 & 3.08 & 15.5 &  \\
PM I19207+0506S & 19:20:46.74 & +05:06:26.5 & -0.060 & -0.149 & 11.45 & 1.34 & PM I19207+0506N & 19:20:46.08 & +05:06:38.5 & -0.058 & -0.140 & 17.42 & 2.96 & 15.4 &  \\
PM I19420+2014S & 19:42:00.86 & +20:14:05.0 & -0.096 & -0.125 & 15.83 & 2.12 & PM I19420+2014N & 19:42:00.88 & +20:14:10.3 & -0.096 & -0.125 & 16.11 & 2.98 & 5.3 & \tablenotemark{m} \\
PM I20072-3519E & 20:07:13.51 & -35:19:50.0 & -0.028 & -0.187 & 15.73 & 1.78 & PM I20072-3519W & 20:07:13.20 & -35:19:52.8 & -0.028 & -0.187 & 16.22 & 3.02 & 4.6 & \tablenotemark{g,k} \\
PM I20343+1151 & 20:34:22.72 & +11:51:59.5 & -0.144 & -0.203 & 12.49 & 1.87 & PM I20343+1152 & 20:34:22.56 & +11:52:03.3 & -0.144 & -0.203 & 15.53 & 3.00 & 4.4 &  \\
NLTT 49474 & 20:34:31.48 & -22:19:24.3 & 0.143 & -0.108 & 11.12 & 1.24 & NLTT 49477 & 20:34:33.49 & -22:17:59.7 & 0.155 & -0.119 & 18.19 & 3.40 & 89.1 &  \\
PM I20487+1406 & 20:48:42.08 & +14:06:59.1 & 0.110 & -0.051 & 14.89 & 1.72 & PM I20487+1407 & 20:48:42.78 & +14:07:01.3 & 0.110 & -0.051 & 19.37 & 3.34 & 10.3 &  \\
PM I21175-4142E & 21:17:32.29 & -41:42:17.3 & 0.013 & -0.157 & 13.89 & 1.74 & PM I21175-4142W & 21:17:31.42 & -41:42:21.2 & 0.013 & -0.157 & 17.64 & 2.84 & 10.5 &  \\
PM I21442+0102N & 21:44:15.64 & +01:02:09.1 & -0.061 & -0.248 & 13.80 & 1.71 & PM I21442+0102S & 21:44:15.92 & +01:02:03.3 & -0.061 & -0.248 & 16.36 & 2.88 & 7.2 &  \\
PM I21536+0010S & 21:53:39.95 & +00:10:20.8 & -0.061 & -0.158 & 14.35 & 2.12 & PM I21536+0010N & 21:53:39.98 & +00:10:37.2 & -0.061 & -0.158 & 19.87 & 3.66 & 16.4 &  \\
NLTT 52532 & 21:57:35.98 & -03:28:09.2 & 0.169 & -0.125 & 14.90 & 1.98 & NLTT 52538 & 21:57:37.93 & -03:28:32.3 & 0.169 & -0.125 & 18.86 & 3.22 & 37.3 & \tablenotemark{m} \\
PM I22296+0620 & 22:29:41.07 & +06:20:02.8 & 0.116 & -0.021 & 13.87 & 2.45 & PM I22297+0620W & 22:29:42.75 & +06:20:09.0 & 0.092 & -0.026 & 18.80 & 3.50 & 25.9 &  \\
PM I22487-5613W & 22:48:44.33 & -56:13:37.0 & 0.145 & -0.067 & 12.59 & 1.60 & PM I22487-5613E & 22:48:45.54 & -56:13:44.0 & 0.145 & -0.067 & 16.67 & 3.41 & 12.3 &  \\
PM I23033-5311 & 23:03:23.47 & -53:11:23.1 & 0.148 & -0.107 & 14.01 & 1.76 & PM I23034-5311 & 23:03:25.79 & -53:11:43.2 & 0.153 & -0.098 & 18.94 & 3.63 & 29.0 &  \\
NLTT 57827 & 23:44:27.89 & -30:55:16.9 & 0.190 & -0.196 & 16.96 & 2.09 & NLTT 57823 & 23:44:24.90 & -30:55:26.2 & 0.190 & -0.201 & 18.17 & 2.77 & 39.5 &  \\
\enddata
\tablenotetext{a}{
Source: PMI stars are binary candidates from LSPM catalog\citep{Lepine2005}; NLTT stars are halo binaries from \cite{Chaname2004}.}
\tablenotetext{b}{
Proper motion (in arcsec/yr) from LSPM catalog for PMI stars and rNLTT catalog for NLTT stars.}
\tablenotetext{c}{
Apparent $V$ magnitude from LSPM catalog and rNLTT catalog; $J$ magnitude from 2MASS catalog\citep{Skrutskie2006}. Note that some of the $V$ magnitudes from LSPM catalog are from a preliminary version of the LSPM-South catalog, and thus might be different from the final version. We selected our sample based on the preliminary version and we therefore list them here.}
\tablenotetext{d}{
Angular separation (in arcsec) between the two components.}
\tablenotetext{e}{
Three possible companions were identified in the preliminary version of the LSPM-South but they are not included in the final version of LSPM catalog, because they are marginal detected on Palomar Sky Survey plates. These secondaries are designated with a ``-2''.
}
\tablenotetext{f}{
Our initial selection found that PM I14475+1134 and PM I14476+1134 are a co-moving pair. However, another star PM I14476+1133 actually has the same proper motion as the other two stars. We therefore think that this might be a triple system.
}
\tablenotetext{g}{
Pair PM I20072-3519E / PM I20072-3519W: PM I20072-3519E was assigned as primary since its $V$ magnitude is brighter than PM I20072-3519W in LSPM catalog. However, our photometry in this work shows that PM I20072-3519W is actually brighter and bluer, so we take PM I20072-3519W as primary and PM I20072-3519E as secondary in Table \ref{table:phot} and Table \ref{table:phot2}.
}
\tablenotetext{h}{
Photometric measurements were not obtained for these four candidate pairs. We therefore didn't include their photometry in Table \ref{table:phot} and Table \ref{table:phot2}.
}
\tablenotetext{i}{
Pair PM I04477-3044W / PM I04477-3044E is the same as NLTT 13968 / NLTT 13970 in \cite{Chaname2004}.
}
\tablenotetext{j}{
Pair PM I12283+1222S / PM I12283+1222N is the same as NLTT 30838 / NLTT 30837 in \cite{Chaname2004}.
}
\tablenotetext{k}{
The primaries of these pairs have $(V-J) > 2.5$ from our measurement and they are very likely to be M+M dwarf pairs.
}
\tablenotetext{l}{
The secondaries of these pairs have $(V-J) < 2.8$ from our measurement and they are very likely to be F/G/K+K dwarf pairs.
}
\tablenotetext{m}{
These pairs have slope $m < 1.5$ or $m > 4$ and therefore are unlikely to be real binaries. See Section 4.2 for more details.
}
\end{deluxetable}
\clearpage
\begin{deluxetable}{lcccccccccccc}
\setlength{\tabcolsep}{0.035in}
\tabletypesize{\scriptsize}
\tablecolumns{13}
\tablewidth{0pc}
\tablecaption{
Nightly Photometric Solutions
\label{table:photsol}
}
\tablehead{
\colhead{} & \colhead{} & \colhead{$B$} & \colhead{} & \colhead{} & \colhead{$V$} & \colhead{} & \colhead{} & \colhead{$R$} & \colhead{} & \colhead{} & \colhead{$I$} &  \colhead{}\\
 \colhead{Date} & \colhead{Zero Point} & \colhead{Extinction} & \colhead{Color} & \colhead{Zero Point} & \colhead{Extinction} & \colhead{Color} & \colhead{Zero Point} & \colhead{Extinction} & \colhead{Color} & \colhead{Zero Point} & \colhead{Extinction} & \colhead{Color}
}
\startdata
2008 Sept 10 & -22.059 & 0.249 & -0.055 & -22.096 & 0.136 & 0.074 & -22.253 & 0.098 & -0.037 & -21.719 & 0.051 & -0.058 \\
2008 Sept 11 & -22.029 & 0.219 & -0.046 & -22.077 & 0.111 & 0.089 & -22.231 & 0.078 & -0.003 & -21.695 & 0.033 & -0.059 \\
2008 Sept 12 & -22.071 & 0.253 & -0.051 & -22.165 & 0.183 & 0.082 & -22.335 & 0.160 & -0.015 & -21.798 & 0.110 & -0.053 \\
2008 Feb 04 & -22.236 & 0.264 & -0.050 & -22.157 & 0.164 & 0.064 & -22.314 & 0.134 & -0.049 & -21.801 & 0.054 & -0.030 \\
2008 Feb 05 & -22.154 & 0.187 & -0.036 & -22.076 & 0.092 & 0.069 & -22.230 & 0.054 & -0.024 & -21.751 & 0.024 & -0.045 \\
2008 Feb 07 & -22.037 & 0.127 & -0.013 & -22.014 & 0.066 & 0.087 & -22.163 & 0.033 & 0.008 & -21.621 & -0.047 & -0.036 \\
2008 Feb 08 & -22.142 & 0.202 & -0.027 & -22.070 & 0.113 & 0.071 & -22.213 & 0.077 & -0.024 & -21.713 & 0.024 & -0.040 \\
\enddata
\end{deluxetable}

\clearpage
\begin{deluxetable}{lccccccccccccc}
\tabletypesize{\scriptsize}
\tablecolumns{14}
\rotate
\tablewidth{0pc}
\tablecaption{
$BVRI$ Photometry of Primaries 
\label{table:phot}
}
\tablehead{
\colhead{Primary ID} & \colhead{$V$} & \colhead{$\sigma(V)$} & \colhead{$B-V$} & \colhead{$\sigma(B-V)$} & \colhead{$V-R$} & \colhead{$\sigma(V-R)$} & \colhead{$V-I$} & \colhead{$\sigma(V-I)$} & \colhead{$J$\tablenotemark{a}} & \colhead{$H$\tablenotemark{a}} & \colhead{$K$\tablenotemark{a}} & \colhead{\# Visits} &\colhead{Notes}
}
\startdata
PM I00025-4644 & 16.501 & 0.023 & 1.182 & 0.029 & 0.757 & 0.024 & 1.335 & 0.028 & 14.228 & 13.595 & 13.451 & 2 &  \\
PM I00329+1805 & 16.326 & 0.023 & 1.389 & 0.031 & 0.891 & 0.025 & 1.682 & 0.028 & 13.641 & 12.970 & 12.764 & 1 &  \\
PM I00422+0731E & 14.878 & 0.023 & 0.837 & 0.027 & 0.527 & 0.024 & 0.994 & 0.028 & 13.169 & 12.663 & 12.581 & 2 &  \\
PM I00592+0705N & 13.861 & 0.023 & 0.665 & 0.027 & 0.373 & 0.024 & 0.737 & 0.028 & 12.577 & 12.239 & 12.179 & 1 &  \\
NLTT 3847 & 15.445 & 0.023 & 1.489 & 0.030 & 1.085 & 0.025 & 2.432 & 0.028 & 11.666 & 11.183 & 10.975 & 1 &  \\
PM I01227+1409 & 10.170 & 0.023 & 0.536 & 0.027 & 0.348 & 0.024 & 0.666 & 0.028 & 9.060 & 8.770 & 8.692 & 1 &  \\
PM I01266-4842W & 14.202 & 0.023 & 1.266 & 0.027 & 0.835 & 0.024 & 1.624 & 0.028 & 11.536 & 10.862 & 10.734 & 1 &  \\
NLTT 4817 & 11.355 & 0.023 & 0.487 & 0.027 & 0.356 & 0.024 & 0.663 & 0.028 & 10.257 & 9.972 & 9.906 & 1 &  \\
PM I01352+0538N & 10.806 & 0.023 & 0.744 & 0.027 & 0.464 & 0.024 & ...... & ...... & 9.260 & 8.793 & 8.698 & 1 & \tablenotemark{b} \\
PM I01430-4959W & 14.467 & 0.023 & 0.786 & 0.027 & 0.508 & 0.024 & 0.913 & 0.028 & 12.866 & 12.378 & 12.337 & 1 &  \\
NLTT  8753 & 16.829 & 0.024 & 1.453 & 0.030 & 0.952 & 0.025 & 1.772 & 0.029 & 13.973 & 13.466 & 13.266 & 1 &  \\
PM I02569-5831N & 13.814 & 0.023 & 1.113 & 0.028 & 0.744 & 0.024 & 1.378 & 0.028 & 11.528 & 10.898 & 10.796 & 1 &  \\
PM I03150+0102 & 10.214 & 0.023 & 0.684 & 0.027 & 0.391 & 0.024 & ...... & ...... & 8.905 & 8.455 & 8.451 & 1 & \tablenotemark{b} \\
PM I03256-3333E & 14.824 & 0.023 & 1.124 & 0.027 & 0.737 & 0.024 & 1.421 & 0.028 & 12.510 & 11.847 & 11.721 & 1 &  \\
NLTT 12296 & 14.641 & 0.023 & 0.976 & 0.027 & 0.658 & 0.024 & ...... & ...... & 12.682 & 12.121 & 12.009 & 1 & \tablenotemark{b} \\
PM I04072+1526N & 10.252 & 0.023 & 0.695 & 0.027 & 0.407 & 0.024 & 0.718 & 0.028 & 8.972 & 8.610 & 8.549 & 1 &  \\
PM I04099+0942E & 14.294 & 0.023 & 0.945 & 0.027 & 0.606 & 0.024 & 1.165 & 0.028 & 12.274 & 11.728 & 11.617 & 1 &  \\
PM I04254-4601 & 14.495 & 0.023 & 0.816 & 0.027 & 0.503 & 0.024 & 0.981 & 0.028 & 12.837 & 12.379 & 12.244 & 1 &  \\
PM I04325-5657N & 12.241 & 0.023 & 0.687 & 0.028 & 0.402 & 0.024 & 0.783 & 0.028 & 10.879 & 10.506 & 10.446 & 1 &  \\
PM I04327+0820 & 13.942 & 0.024 & 0.617 & 0.028 & 0.490 & 0.025 & 0.873 & 0.029 & 12.360 & 11.973 & 11.867 & 1 &  \\
PM I04332+0013 & 11.432 & 0.023 & 0.646 & 0.027 & 0.368 & 0.024 & 0.705 & 0.028 & 10.207 & 9.854 & 9.778 & 1 &  \\
PM I04477-3044 & 11.619 & 0.023 & 0.632 & 0.027 & 0.414 & 0.024 & 0.810 & 0.028 & 10.277 & 9.922 & 9.898 & 2 &  \\
NLTT 14407 & 11.661 & 0.023 & 1.114 & 0.027 & 0.686 & 0.024 & 1.252 & 0.028 & 9.611 & 9.037 & 8.888 & 1 &  \\
PM I05137+0647W & 12.288 & 0.024 & 0.629 & 0.029 & 0.484 & 0.025 & 0.846 & 0.029 & 10.893 & 10.536 & 10.496 & 1 &  \\
PM I05195+0903E & 14.207 & 0.023 & 0.830 & 0.027 & 0.523 & 0.024 & 1.023 & 0.028 & 12.415 & 12.021 & 11.912 & 1 &  \\
PM I05484-3617Nn & 14.607 & 0.023 & 0.827 & 0.027 & 0.505 & 0.024 & 0.973 & 0.028 & 12.986 & 12.520 & 12.391 & 1 &  \\
PM I06032+1921N & 9.289 & 0.023 & 0.580 & 0.027 & ...... & ...... & ...... & ...... & 8.001 & 7.695 & 7.583 & 1 & \tablenotemark{b} \\
PM I06050+0723S & 15.433 & 0.023 & 1.392 & 0.027 & 0.903 & 0.024 & 1.682 & 0.028 & 12.767 & 12.169 & 11.959 & 2 &  \\
PM I06394-3030E & 15.275 & 0.023 & 1.414 & 0.027 & 0.908 & 0.024 & 1.734 & 0.028 & 12.572 & 12.062 & 11.818 & 2 &  \\
PM I06436+0851 & 14.124 & 0.023 & 1.194 & 0.028 & 0.852 & 0.026 & 1.472 & 0.029 & 11.701 & 11.110 & 11.001 & 2 &  \\
PM I08152-6337 & 12.084 & 0.023 & 1.139 & 0.027 & 0.727 & 0.024 & 1.303 & 0.028 & 9.957 & 9.366 & 9.248 & 4 &  \\
PM I08239-7549W & 10.960 & 0.023 & 0.833 & 0.027 & 0.479 & 0.024 & 0.863 & 0.028 & 9.534 & 9.078 & 8.992 & 1 &  \\
PM I08386-3856 & 12.384 & 0.023 & 0.889 & 0.028 & 0.538 & 0.024 & 1.026 & 0.028 & 10.727 & 10.209 & 10.146 & 1 &  \\
PM I09502+0509E & 11.673 & 0.023 & 0.805 & 0.027 & 0.474 & 0.024 & 0.923 & 0.028 & 10.195 & 9.706 & 9.615 & 1 &  \\
PM I10105+1203W & 12.554 & 0.023 & 0.785 & 0.027 & 0.470 & 0.024 & 0.917 & 0.028 & 10.975 & 10.488 & 10.418 & 2 &  \\
PM I10520+1521N & 16.024 & 0.023 & 1.115 & 0.027 & 0.725 & 0.024 & 1.338 & 0.028 & 13.862 & 13.281 & 13.117 & 2 &  \\
PM I11110-4414 & 12.841 & 0.023 & 0.723 & 0.027 & 0.439 & 0.024 & 0.857 & 0.028 & 11.409 & 11.009 & 10.891 & 2 &  \\
PM I11125-3512 & 15.125 & 0.023 & 1.035 & 0.027 & 0.629 & 0.024 & 1.195 & 0.028 & 13.099 & 12.523 & 12.433 & 2 &  \\
NLTT 27188 & 12.550 & 0.023 & 0.538 & 0.027 & 0.343 & 0.024 & 0.697 & 0.028 & 11.392 & 11.054 & 11.008 & 1 &  \\
PM I11263+2047Ee & 13.453 & 0.023 & 1.155 & 0.027 & 0.762 & 0.024 & 1.400 & 0.028 & 11.164 & 10.534 & 10.404 & 2 &  \\
PM I11330+1318N & 9.400 & 0.023 & 0.621 & 0.027 & 0.380 & 0.024 & ...... & ...... & 8.281 & 7.976 & 7.895 & 1 & \tablenotemark{b} \\
PM I11392-4118N & 13.244 & 0.023 & 1.424 & 0.027 & 0.929 & 0.024 & 1.847 & 0.028 & 10.354 & 9.713 & 9.513 & 2 &  \\
PM I11584-4155E & ...... & ...... & ...... & ...... & ...... & ...... & ...... & ...... & 7.330 & 6.878 & 6.812 & 0 & \tablenotemark{b} \\
PM I12170+0742E & 15.055 & 0.023 & 0.759 & 0.027 & 0.475 & 0.024 & 0.899 & 0.028 & 13.550 & 13.046 & 13.025 & 2 &  \\
PM I12237+0625 & 13.802 & 0.024 & 0.979 & 0.028 & 0.649 & 0.025 & 1.185 & 0.028 & 11.864 & 11.213 & 11.132 & 2 &  \\
PM I12277+1334 & 15.947 & 0.023 & 1.102 & 0.028 & 0.719 & 0.024 & 1.330 & 0.028 & 13.779 & 13.199 & 13.000 & 1 &  \\
PM I12283+1222S & ...... & ...... & ...... & ...... & ...... & ...... & ...... & ...... & 11.398 & 11.115 & 11.073 & 0 & \tablenotemark{b} \\
PM I12440+0625E & 12.937 & 0.023 & 0.454 & 0.028 & 0.305 & 0.025 & 0.621 & 0.028 & 11.890 & 11.630 & 11.555 & 1 &  \\
PM I12508+0757 & 10.666 & 0.023 & 0.567 & 0.027 & 0.346 & 0.024 & 0.690 & 0.028 & 9.572 & 9.256 & 9.179 & 2 &  \\
PM I13116+1106 & 12.933 & 0.023 & 0.518 & 0.028 & 0.352 & 0.025 & 0.678 & 0.028 & 11.830 & 11.507 & 11.451 & 1 &  \\
NLTT 33282 & 15.404 & 0.023 & 1.331 & 0.028 & 0.811 & 0.025 & 1.547 & 0.028 & 12.920 & 12.295 & 12.093 & 2 &  \\
PM I13133-4153N & 13.529 & 0.023 & 0.970 & 0.028 & 0.624 & 0.024 & 1.176 & 0.028 & 11.578 & 10.954 & 10.892 & 2 &  \\
PM I13167+0810E & 11.062 & 0.023 & 0.847 & 0.028 & 0.509 & 0.024 & 0.948 & 0.028 & 9.498 & 9.002 & 8.904 & 1 &  \\
PM I13372-4244E & 17.200 & 0.027 & 1.368 & 0.062 & 0.838 & 0.028 & 1.483 & 0.032 & 14.841 & 14.186 & 13.945 & 1 &  \\
PM I14055+0244S & 15.479 & 0.023 & 1.039 & 0.027 & 0.689 & 0.024 & 1.237 & 0.028 & 13.409 & 12.858 & 12.784 & 1 &  \\
PM I14124+0517S & 13.275 & 0.023 & 0.717 & 0.028 & 0.414 & 0.025 & 0.825 & 0.028 & 11.902 & 11.496 & 11.424 & 1 &  \\
PM I14136-3634E & 15.190 & 0.023 & 1.037 & 0.027 & 0.645 & 0.024 & 1.236 & 0.028 & 13.152 & 12.590 & 12.430 & 1 &  \\
PM I14475+1134 & 12.792 & 0.023 & 0.741 & 0.028 & 0.439 & 0.025 & 0.836 & 0.028 & 11.375 & 10.988 & 10.908 & 1 &  \\
PM I15413+1349N & 15.145 & 0.024 & 1.474 & 0.032 & 0.944 & 0.025 & 1.919 & 0.029 & 12.105 & 11.549 & 11.325 & 1 &  \\
PM I16008+0146E & 13.262 & 0.023 & 0.881 & 0.027 & 0.665 & 0.024 & 1.170 & 0.028 & 11.272 & 10.719 & 10.597 & 2 &  \\
PM I16519-4806N & 15.420 & 0.023 & 1.077 & 0.027 & 0.729 & 0.024 & 1.337 & 0.028 & 13.094 & 12.551 & 12.409 & 2 &  \\
PM I17135+1909 & 11.051 & 0.023 & 0.759 & 0.027 & 0.497 & 0.024 & 0.882 & 0.028 & 9.593 & 9.163 & 9.107 & 1 &  \\
PM I19207+0506S & 11.336 & 0.023 & 0.582 & 0.027 & 0.368 & 0.024 & 0.704 & 0.028 & 10.110 & 9.801 & 9.767 & 2 &  \\
PM I19420+2014S & 16.621 & 0.024 & 1.385 & 0.034 & 0.927 & 0.026 & 1.809 & 0.029 & 13.714 & 13.081 & 12.882 & 2 &  \\
PM I20072-3519W & 17.011 & 0.027 & 1.541 & 0.045 & 1.080 & 0.029 & 2.426 & 0.031 & 13.199 & 12.636 & 12.398 & 2 &  \\
PM I20343+1151 & 12.300 & 0.023 & 0.884 & 0.027 & 0.539 & 0.024 & 0.977 & 0.028 & 10.624 & 10.141 & 10.044 & 2 &  \\
NLTT 49474 & 11.047 & 0.023 & 0.573 & 0.027 & 0.352 & 0.024 & 0.691 & 0.028 & 9.882 & 9.541 & 9.519 & 2 &  \\
PM I20487+1406 & 15.420 & 0.023 & 1.151 & 0.027 & 0.740 & 0.024 & 1.367 & 0.028 & 13.172 & 12.565 & 12.399 & 2 &  \\
PM I21175-4142E & 13.620 & 0.023 & 0.731 & 0.027 & 0.415 & 0.024 & 0.851 & 0.028 & 12.148 & 11.779 & 11.684 & 2 &  \\
PM I21442+0102N & 14.468 & 0.023 & 1.199 & 0.027 & 0.782 & 0.024 & 1.434 & 0.028 & 12.091 & 11.452 & 11.323 & 2 &  \\
PM I21536+0010S & 14.032 & 0.023 & 0.838 & 0.027 & 0.536 & 0.024 & 1.051 & 0.028 & 12.234 & 11.776 & 11.639 & 2 &  \\
NLTT 52532 & 15.538 & 0.023 & 1.360 & 0.028 & 0.863 & 0.024 & 1.609 & 0.028 & 12.930 & 12.383 & 12.179 & 2 &  \\
PM I22296+0620 & 13.716 & 0.023 & 1.118 & 0.027 & 0.715 & 0.024 & 1.351 & 0.028 & 11.424 & 10.851 & 10.707 & 2 &  \\
PM I22487-5613W & 12.700 & 0.023 & 0.902 & 0.028 & 0.541 & 0.024 & 0.978 & 0.028 & 10.989 & 10.511 & 10.417 & 2 &  \\
PM I23033-5311 & 13.389 & 0.023 & 0.522 & 0.027 & 0.331 & 0.024 & 0.662 & 0.028 & 12.249 & 11.911 & 11.848 & 2 &  \\
NLTT 57827 & 17.154 & 0.024 & 1.166 & 0.031 & 0.735 & 0.025 & 1.367 & 0.029 & 14.887 & 14.385 & 14.179 & 2 &  \\
\enddata
\tablenotetext{a}{
JHK photometry are from 2MASS catalog \citep{Skrutskie2006}.}
\tablenotetext{b}{
Photometry was not measured for one or more colors.
}
\end{deluxetable}

\clearpage
\begin{deluxetable}{lccccccccccccc}
\tabletypesize{\scriptsize}
\tablecolumns{14}
\rotate
\tablewidth{0pc}
\tablecaption{
$BVRI$ Photometry of Secondaries
\label{table:phot2}
}
\tablehead{
\colhead{Secondary ID} & \colhead{$V$} & \colhead{$\sigma(V)$} & \colhead{$B-V$} & \colhead{$\sigma(B-V)$} & \colhead{$V-R$} & \colhead{$\sigma(V-R)$} & \colhead{$V-I$} & \colhead{$\sigma(V-I)$} & \colhead{$J$\tablenotemark{a}} & \colhead{$H$\tablenotemark{a}} & \colhead{$K$\tablenotemark{a}} & \colhead{\# Visits} &\colhead{Notes}
}
\startdata
PM I00026-4644 & 20.349 & 0.148 & 2.132 & 0.909 & 0.955 & 0.151 & 2.098 & 0.143 & 16.952 & 16.173 & 15.917 & 2 \\
PM I00329+1805-2 & 20.933 & 0.247 & 1.557 & 0.961 & 1.235 & 0.248 & 2.993 & 0.236 & 16.156 & 16.016 & 15.424 & 1 \\
PM I00422+0731W & 18.975 & 0.046 & 1.630 & 0.178 & 0.919 & 0.048 & 1.880 & 0.048 & 15.922 & 15.524 & 14.923 & 2 \\
PM I00592+0705S & 19.367 & 0.074 & 1.758 & 0.302 & 1.133 & 0.075 & 2.275 & 0.073 & 15.753 & 15.332 & 15.076 & 1 \\
NLTT 3849 & 15.639 & 0.024 & 1.484 & 0.031 & 1.114 & 0.025 & 2.509 & 0.029 & 11.752 & 11.275 & 11.052 & 1 \\
PM I01226+1409E & 19.225 & 0.047 & 2.000 & 0.227 & 1.239 & 0.048 & 2.777 & 0.049 & 15.006 & 14.563 & 14.169 & 1 \\
PM I01266-4842E & 18.723 & 0.030 & 1.632 & 0.083 & 1.232 & 0.033 & 2.784 & 0.036 & 14.432 & 13.938 & 13.701 & 1 \\
NLTT 4814 & 16.518 & 0.025 & 1.405 & 0.037 & 0.953 & 0.027 & 1.776 & 0.030 & 13.712 & 13.157 & 12.938 & 1 \\
PM I01352+0538S & 17.082 & 0.029 & 1.542 & 0.047 & 1.150 & 0.031 & 2.654 & 0.033 & 12.968 & 12.479 & 12.271 & 1 \\
PM I01430-4959E & 18.072 & 0.025 & 1.425 & 0.040 & 0.956 & 0.027 & 1.784 & 0.031 & 15.258 & 14.767 & 14.532 & 1 \\
NLTT 8759 & 17.665 & 0.025 & 1.627 & 0.037 & 1.070 & 0.026 & 2.080 & 0.030 & 14.434 & 13.881 & 13.673 & 1 \\
PM I02569-5831S & 20.879 & 0.100 & 2.236 & 0.505 & 1.406 & 0.100 & 3.327 & 0.097 & 15.989 & 15.195 & 15.118 & 1 \\
PM I03150+0103 & 14.669 & 0.023 & 1.483 & 0.028 & 0.930 & 0.025 & 1.958 & 0.029 & 11.622 & 11.043 & 10.855 & 1 \\
PM I03256-3333Wn & 16.290 & 0.023 & 1.351 & 0.028 & 0.864 & 0.024 & 1.658 & 0.029 & 13.639 & 13.010 & 12.845 & 1 \\
NLTT 12294 & 18.503 & 0.027 & 1.558 & 0.048 & 1.053 & 0.029 & 2.045 & 0.032 & 15.260 & 14.786 & 14.518 & 1 \\
PM I04072+1526S & 18.828 & 0.030 & 1.723 & 0.062 & 1.309 & 0.033 & 3.033 & 0.034 & 14.260 & 13.693 & 13.449 & 2 \\
PM I04099+0942W & 15.608 & 0.024 & 1.263 & 0.028 & 0.726 & 0.025 & 1.428 & 0.029 & 13.170 & 12.531 & 12.351 & 1 \\
PM I04255-4601 & 19.477 & 0.036 & 1.390 & 0.081 & 0.967 & 0.040 & 2.005 & 0.044 & 16.408 & 15.911 & 15.317 & 1 \\
PM I04325-5657S & 16.867 & 0.024 & 1.452 & 0.030 & 0.970 & 0.026 & 2.062 & 0.029 & 13.650 & 13.179 & 12.888 & 1 \\
PM I04327+0820-2 & 20.259 & 0.150 & 1.201 & 0.495 & 1.194 & 0.153 & 2.248 & 0.146 & 16.796 & 16.097 & 15.993 & 1 \\
PM I04333+0014 & 16.513 & 0.024 & 1.565 & 0.030 & 0.985 & 0.026 & 2.042 & 0.029 & 13.336 & 12.779 & 12.580 & 1 \\
PM I04477-3044E & 18.775 & 0.033 & 1.582 & 0.074 & 1.165 & 0.037 & 2.591 & 0.038 & 15.041 & 14.533 & 14.353 & 1 \\
NLTT 14408 & 14.855 & 0.024 & 1.524 & 0.030 & 1.015 & 0.025 & 2.093 & 0.029 & 11.646 & 11.143 & 10.939 & 1 \\
PM I05137+0647E & 19.132 & 0.049 & 1.655 & 0.105 & 1.228 & 0.051 & 2.637 & 0.048 & 15.161 & 14.606 & 14.292 & 1 \\
PM I05195+0903W & 17.861 & 0.030 & 1.569 & 0.081 & 0.961 & 0.031 & 1.829 & 0.034 & 14.879 & 14.330 & 14.030 & 1 \\
PM I05484-3617S & 18.465 & 0.027 & 1.517 & 0.047 & 0.975 & 0.029 & 1.928 & 0.033 & 15.488 & 15.020 & 14.674 & 1 \\
PM I06032+1921S & 13.089 & 0.024 & 1.244 & 0.029 & 0.890 & 0.025 & 1.599 & 0.028 & 10.587 & 10.011 & 9.822 & 2 \\
PM I06050+0723N & 16.994 & 0.024 & 1.592 & 0.030 & 1.003 & 0.026 & 2.006 & 0.029 & 13.838 & 13.293 & 13.084 & 2 \\
PM I06394-3030W & 16.664 & 0.024 & 1.545 & 0.029 & 0.984 & 0.025 & 1.911 & 0.029 & 13.688 & 13.209 & 12.998 & 2 \\
PM I06436+0851-2 & 19.698 & 0.035 & 1.574 & 0.083 & 1.342 & 0.038 & 3.116 & 0.039 & 15.129 & 14.593 & 14.396 & 2 \\
PM I08153-6337 & 17.364 & 0.024 & 1.701 & 0.029 & 1.209 & 0.025 & 2.708 & 0.028 & 13.237 & 12.672 & 12.402 & 4 \\
PM I08239-7549E & 17.811 & 0.027 & 1.699 & 0.134 & 1.303 & 0.119 & 3.029 & 0.118 & 13.227 & 12.676 & 12.388 & 1 \\
PM I08386-3857 & 17.643 & 0.024 & 1.597 & 0.031 & 1.125 & 0.025 & 2.514 & 0.029 & 13.845 & 13.349 & 13.161 & 2 \\
PM I09502+0509W & 18.590 & 0.029 & 1.724 & 0.060 & 1.215 & 0.033 & 2.774 & 0.034 & 14.566 & 13.995 & 13.811 & 1 \\
PM I10105+1203E & 17.849 & 0.025 & 1.542 & 0.125 & 1.093 & 0.113 & 2.267 & 0.114 & 14.374 & 13.793 & 13.633 & 1 \\
PM I10520+1521S & 17.947 & 0.024 & 1.447 & 0.032 & 0.932 & 0.026 & 1.813 & 0.030 & 15.123 & 14.547 & 14.438 & 2 \\
PM I11109-4416 & 16.115 & 0.024 & 1.402 & 0.028 & 0.889 & 0.025 & 1.686 & 0.029 & 13.413 & 12.879 & 12.661 & 2 \\
PM I11124-3512 & 18.798 & 0.025 & 1.664 & 0.043 & 1.080 & 0.027 & 2.318 & 0.031 & 15.222 & 14.643 & 14.368 & 2 \\
NLTT 27182 & 14.669 & 0.023 & 1.026 & 0.027 & 0.660 & 0.024 & 1.285 & 0.028 & 12.521 & 11.943 & 11.824 & 1 \\
PM I11263+2047Ew & 17.377 & 0.024 & 1.594 & 0.031 & 1.079 & 0.026 & 2.393 & 0.030 & 13.762 & 13.267 & 13.035 & 2 \\
PM I11330+1318S & 16.879 & 0.025 & 1.595 & 0.033 & 1.192 & 0.026 & 2.728 & 0.030 & 12.791 & 12.184 & 11.917 & 1 \\
PM I11392-4118S & 15.556 & 0.024 & 1.559 & 0.029 & 1.108 & 0.025 & 2.470 & 0.029 & 11.783 & 11.164 & 10.919 & 2 \\
PM I11584-4155W & 16.042 & 0.024 & 1.678 & 0.032 & 1.212 & 0.026 & 2.819 & 0.029 & 11.777 & 11.293 & 11.038 & 2 \\
PM I12170+0742W & 18.080 & 0.024 & 1.337 & 0.032 & 0.859 & 0.026 & 1.586 & 0.030 & 15.418 & 14.929 & 14.886 & 2 \\
PM I12237+0624 & 18.252 & 0.024 & 1.578 & 0.036 & 1.081 & 0.026 & 2.386 & 0.030 & 14.631 & 14.166 & 13.799 & 2 \\
PM I12277+1336 & 18.156 & 0.025 & 1.452 & 0.038 & 0.929 & 0.027 & 1.744 & 0.030 & 15.495 & 14.808 & 14.643 & 1 \\
PM I12283+1222N & 19.200 & 0.030 & 1.583 & 0.060 & 1.045 & 0.033 & 2.047 & 0.036 & 16.024 & 15.374 & 15.592 & 1 \\
PM I12440+0625We & 19.186 & 0.033 & 1.435 & 0.069 & 0.949 & 0.037 & 1.831 & 0.040 & 16.131 & 15.624 & 15.476 & 1 \\
PM I12507+0758 & 18.105 & 0.025 & 1.636 & 0.036 & 1.134 & 0.026 & 2.544 & 0.030 & 14.250 & 13.738 & 13.511 & 2 \\
PM I13116+1105 & 17.515 & 0.024 & 1.321 & 0.032 & 0.864 & 0.026 & 1.536 & 0.030 & 15.028 & 14.479 & 14.303 & 1 \\
NLTT 33283 & 15.872 & 0.024 & 1.441 & 0.029 & 0.864 & 0.025 & 1.657 & 0.029 & 13.228 & 12.633 & 12.428 & 2 \\
PM I13133-4153S & 18.268 & 0.024 & 1.695 & 0.037 & 1.076 & 0.026 & 2.340 & 0.030 & 14.701 & 14.204 & 13.979 & 2 \\
PM I13167+0810W & 19.231 & 0.049 & 1.853 & 0.129 & 1.405 & 0.059 & 3.245 & 0.051 & 14.393 & 13.863 & 13.570 & 1 \\
PM I13372-4244W & 19.566 & 0.127 & 1.757 & 0.661 & 1.135 & 0.129 & 1.964 & 0.127 & 16.120 & 15.880 & 15.630 & 1 \\
PM I14055+0244N & 20.128 & 0.088 & 1.880 & 0.298 & 1.126 & 0.102 & 2.466 & 0.091 & 16.502 & 16.379 & 15.847 & 1 \\
PM I14124+0517N & 19.399 & 0.036 & 1.409 & 0.077 & 1.135 & 0.039 & 2.394 & 0.041 & 15.786 & 15.347 & 15.170 & 1 \\
PM I14136-3634W & 18.033 & 0.025 & 1.463 & 0.039 & 0.963 & 0.027 & 1.874 & 0.030 & 15.225 & 14.559 & 14.405 & 1 \\
PM I14476+1133 & 19.746 & 0.058 & 1.847 & 0.192 & 0.943 & 0.066 & 1.924 & 0.073 & 17.060 & 16.197 & 16.131 & 1 \\
PM I15413+1349S & 17.286 & 0.035 & 1.267 & 0.083 & 1.308 & 0.037 & 2.419 & 0.038 & 13.572 & 13.014 & 12.833 & 1 \\
PM I16008+0146W & 17.897 & 0.026 & 1.653 & 0.059 & 1.109 & 0.027 & 2.260 & 0.030 & 14.383 & 13.788 & 13.656 & 2 \\
PM I16519-4806S & 18.411 & 0.035 & 1.820 & 0.087 & 1.173 & 0.038 & 2.158 & 0.040 & 14.220 & 13.191 & 12.770 & 2 \\
PM I17134+1910 & 16.529 & 0.025 & 1.579 & 0.040 & 1.100 & 0.027 & 2.447 & 0.030 & 12.715 & 12.154 & 11.969 & 2 \\
PM I19207+0506N & 18.018 & 0.029 & 1.697 & 0.063 & 1.086 & 0.031 & 2.261 & 0.033 & 14.456 & 13.915 & 13.585 & 2 \\
PM I19420+2014N & 16.883 & 0.025 & 1.562 & 0.040 & 1.111 & 0.026 & 2.397 & 0.030 & 13.132 & 12.519 & 12.313 & 2 \\
PM I20072-3519E & 18.158 & 0.045 & 1.616 & 0.106 & 1.198 & 0.048 & 2.680 & 0.047 & 13.950 & 13.411 & 13.199 & 2 \\
PM I20343+1152 & 16.092 & 0.024 & 1.188 & 0.111 & 1.287 & 0.113 & 2.328 & 0.113 & 12.535 & 11.939 & 11.702 & 1 \\
NLTT 49477 & 18.601 & 0.052 & 1.984 & 0.211 & 1.113 & 0.055 & 2.502 & 0.054 & 14.789 & 14.222 & 14.027 & 2 \\
PM I20487+1407 & 19.405 & 0.054 & 1.693 & 0.210 & 1.117 & 0.057 & 2.224 & 0.056 & 16.026 & 15.475 & 15.054 & 2 \\
PM I21175-4142W & 18.069 & 0.029 & 1.435 & 0.064 & 0.997 & 0.030 & 2.097 & 0.033 & 14.804 & 14.223 & 13.949 & 2 \\
PM I21442+0102S & 16.666 & 0.026 & 1.494 & 0.037 & 0.973 & 0.027 & 2.032 & 0.030 & 13.476 & 12.921 & 12.715 & 2 \\
PM I21536+0010N & 19.532 & 0.084 & 1.576 & 0.298 & 1.190 & 0.100 & 2.214 & 0.108 & 16.206 & 15.621 & 15.374 & 2 \\
NLTT 52538 & 19.027 & 0.050 & 1.761 & 0.201 & 1.065 & 0.052 & 2.136 & 0.050 & 15.693 & 15.357 & 14.804 & 2 \\
PM I22297+0620W & 20.068 & 0.272 & 1.916 & 1.912 & 1.426 & 0.291 & 3.107 & 0.261 & 15.303 & 14.775 & 14.437 & 2 \\
PM I22487-5613E & 16.783 & 0.024 & 1.501 & 0.031 & 1.018 & 0.025 & 2.223 & 0.029 & 13.255 & 12.685 & 12.442 & 2 \\
PM I23034-5311 & 18.510 & 0.032 & 1.439 & 0.080 & 1.016 & 0.033 & 2.028 & 0.036 & 15.305 & 14.810 & 14.499 & 2 \\
NLTT 57823 & 18.269 & 0.027 & 1.441 & 0.054 & 0.892 & 0.029 & 1.665 & 0.032 & 15.421 & 15.064 & 14.749 & 2 \\
\enddata
\tablenotetext{a}{
JHK photometry are from 2MASS catalog \citep{Skrutskie2006}.}

\end{deluxetable}

\end{document}